\documentclass[twocolumn]{aastex62}
\usepackage{amsmath,amsfonts,amssymb}
\usepackage{hhline}
\usepackage{braket}

\definecolor{dred}{rgb}{0.64,0.08,0.18}
\definecolor{dblue}{rgb}{0,0.45,0.74}
\definecolor{dpurp}{rgb}{0.2539,0.2653,0.53}
\definecolor{dgreen}{rgb}{0.1636,0.4711,0.5581}

\graphicspath{{./}{Figures/}}
\shortauthors{Bechter A. et al.}

\begin{document}

\title{Characterization of Single-Mode Fiber Coupling at the Large Binocular Telescope}

\correspondingauthor{Andrew Bechter}
\email{abechter@nd.ed}

\author{Andrew J. Bechter}
\affiliation{University of Notre Dame, Department of Physics, 225 Nieuwland Science Hall, Notre Dame, IN 46556, USA}

\author{Jonathan Crass}
\affiliation{University of Notre Dame, Department of Physics, 225 Nieuwland Science Hall, Notre Dame, IN 46556, USA}

\author{Jonathan Tesch}
\affiliation{Jet Propulsion Laboratory, California Institute of Technology, 4800 Oak Grove Drive, Pasadena, CA 91109, USA}

\author{Justin R. Crepp}
\affiliation{University of Notre Dame, Department of Physics, 225 Nieuwland Science Hall, Notre Dame, IN 46556, USA}

\author{Eric B. Bechter}
\affiliation{University of Notre Dame, Department of Physics, 225 Nieuwland Science Hall, Notre Dame, IN 46556, USA}

\begin{abstract}
    Optimizing on-sky single-mode fiber (SMF) injection is an essential part of developing precise Doppler spectrometers and new astrophotonics technologies. We installed and tested a prototype SMF injection system at the Large Binocular Telescope (LBT) in April 2016. The fiber injection unit was built as part of the de-risking process for a new instrument named iLocater that will use adaptive optics (AO) to feed a high resolution, near-infrared spectrograph. In this paper we report Y-band SMF coupling measurements for bright, M-type stars. We compare theoretical expectations for delivered Strehl ratio and SMF coupling to experimental results, and evaluate fundamental effects that limit injection efficiency. We find the pupil geometry of the telescope itself limits fiber coupling to a maximum efficiency of $\rho_{\rm tel}\approx 0.78$. Further analysis shows the individual impact of AO correction, tip-tilt residuals, and static (non-common-path) aberrations contribute coupling coefficients of $\rho_{\rm Strehl}\approx0.33$, $\rho_{\rm tip/tilt}\approx0.84$, and $\rho_{\rm ncpa}\approx0.8$ respectively. Combined, these effects resulted in an average Y-band SMF efficiency of $0.18$ for all observations. Finally, we investigate the impact of fiber coupling on radial velocity (RV) precision as a function of stellar apparent magnitude. 
\end{abstract}

\keywords{Single-mode fibers, Adaptive optics, Doppler Spectroscopy, Astrophotonics}

\section{Introduction} 
Adaptive optics (AO) correction translates directly to increased spectral resolution, allowing spectrographs on large telescopes to highly sample stellar absorption lines and monitor the evolution of wavelength-dependent line asymmetries~\citep{Crepp2014}. Diffraction-limited observations permit the use of a single-mode fiber (SMF), which transmits only a single propagation mode and translates into several benefits for RV instruments: 1) Stabilizing the instrument point spread function (PSF) and eliminating modal noise entirely~\citep{Snyder1983,Robertson2012}; 2) The small core diameter of the SMF enables ultra-high spectral resolution ($R>200k$) with a compact diffraction grating, effectively miniaturizing the spectrograph~\citep{Jovanovic2016}; 3) High spectral resolution permits the measurement and possible correction of stellar noise~\citep{Davis2017} ; and 4) High spatial resolution from AO systems allows for the confirmation of (transiting) planets in multiple stars, in particular which star(s) hosts the planet(s)~\citep{Ciardi2015}.

However, with the introduction of a SMF, there are new spectrograph calibration effects to consider. In particular, birefringence is a property of SMFs that can result in variable polarization states propagating through the fiber. This effect, outlined in \citet{Halverson2015}, scales with the degree of polarization ranging from 10 cm/s for weakly polarized light ($\approx$1\%) to several m/s for strongly polarized light ($\approx$100\%). Characterization of birefringence in SMF feeds and mitigating the impact on RV is an on-going area of research (Bechter et al. in prep)

With a core size of only $\approx$~5-10$\; \mu$m,\footnote{For comparison, multi-mode fibers generally have core diameters in the range of $d=50-200\;\mu$m for astronomical applications.} injecting starlight from a telescope directly into SMF's is non-trivial, requiring exquisite beam control~\citep{Serabyn2010,Jovanovic2014}. Developing an efficient SMF injection system thus represents a critical step in realizing the potential of a diffraction-limited spectrograph capable of extremely precise Doppler RV measurements~\citep{Bechter2015,Blake2015,Bechter2016,Bechter2019}.

SMFs also show promise to be beneficial in non-traditional astronomical instruments. The small core diameter enables high-spatial resolution filtering of the input field, a particularly important feature for high dispersion coronagraphy, exoplanet atmospheric characterization, and fiber-nulling \citep{Mawet2017,Wang2017}. In addition, single-mode devices are an integral part of the emerging astro-photonics field, including fiber-based wavefront sensing and photonic reformatters \citep{Bland-Hawthorn2010,Hottinger2018,Harris2018}.

The challenge for all SMF-fed astronomical instruments is to efficiently inject starlight into the fiber. In order to couple and propagate light through a SMF with minimal loss, the incident complex electric field from a telescope must be precisely matched to the boundary conditions of the SMF. Specifically, achieving high coupling efficiency depends on the properties of the incident complex electric field in addition to the intensity field distribution of the fiber acceptance mode~\citep{Shaklan1988,Ruilier2001}. These mode-matching conditions dictate an optimal incident beam must have: a flat wave front, Gaussian amplitude distribution matching the fiber mode shape, and a beam angle that does not exceed the fiber diffraction angle. Deviations from these conditions degrade coupling efficiency into the fiber. On a ground-based telescope, an AO system is essential for matching the PSF to the fiber by correcting wavefront errors caused by atmospheric distortions in the input electric field. Moreover, Phase-Induced Amplitude Apodization (PIAA) optics have been shown to further optimize mode-matching by remapping diffraction effects from the telescope secondary mirror and support structures to reshape the telescope PSF with a more Gaussian-like amplitude distribution~\citep{Guyon2003,Jovanovic2016a}.

A forerunner SMF-injection system has been under development for ``iLocater": a diffraction-limited RV spectrometer that utilizes AO to inject starlight into SMF's at the Large Binocular Telescope (LBT) \citep{Crepp2016}. In April 2016, six half-nights of observations were scheduled to perform initial tests with a SMF injection prototype called the ``iLocater Demonstrator." The demonstration system was installed on the DX side (right side) of the LBT and uses one of the LBTI pyramid wavefront sensors\citep{Esposito2010}. 

\begin{figure}
    \centering
    \includegraphics[width = 3.3in]{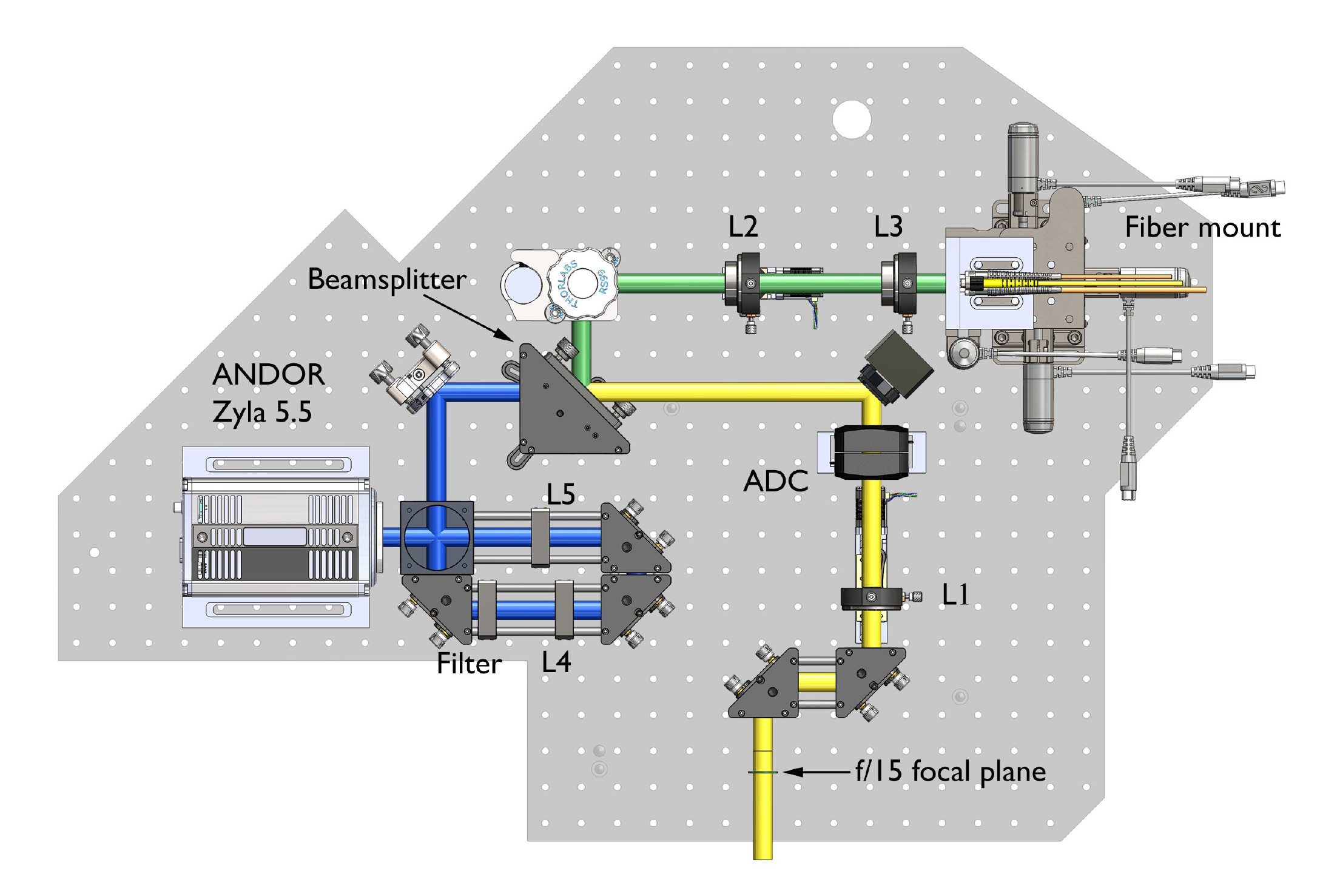}
    \caption{A simplified cartoon of the AC demonstration system showing only the essential optical components. All wavelengths (yellow) propagate initially through the optical train. A beamsplitter directs Y- and J-band light toward the fiber arm (green). Shorter wavelengths are transmitted through the beamsplitter to the imaging arm.}
    \label{fig:block}
\end{figure}

The demonstrator system consisted of three optical channels shown in Figure~\ref{fig:block}; Light from the telescope was first collimated at the entrance of the optical system, referred to as the ``common optics channel." This channel is partitioned using collimated light and a dichroic beamsplitter into a ``fiber coupling channel" and ``imaging channel." The fiber coupling channel consists of a pair of lenses for fiber injection and retractable pick-off mirror to measure the incident optical power in the Y-band. The imaging channel was used to monitor the PSF in the I-band. Details of the demonstrator optical and mechanical design as well as initial commissioning measurements can be found in \citet{Bechter2015} and \citet{Bechter2016}, respectively. 

In this paper we characterize the spatial and temporal performance of SMF coupling efficiency and the LBT AO system using on-sky observations and day-time engineering tests. We describe the methodology used to measure SMF coupling efficiency and reduce additional diagnostic image data in \S\ref{sec:procedure}. We provide a summary of observations in \S\ref{sec:summary}. On-sky SMF coupling efficiency is presented in \S\ref{sec:SMFcoupling}. Relationships between fiber coupling efficiency, telescope pupil geometry, AO correction and tip-tilt residuals are derived in \S\ref{sec:Pupil}, \S\ref{sec:Strehl}, and \S\ref{sec:TT} respectively. We further estimate the impact of static aberrations and demonstrate a first step for mitigation using phase retrieval in \S\ref{sec:static}. Finally we translate fiber coupling losses to RV uncertainty and estimate limiting magnitudes in \S\ref{sec:RV}.

\begin{deluxetable*}{cccccccccc}
    \tablecaption{Observing Summary\label{tab:obs_sum}}
	\tablehead{	
	\colhead{ID} & \colhead{Sp. type} & \colhead{V} & \colhead{J} & \colhead{AO modes} & \colhead{Seeing ('')} & \colhead{Airmass} & \colhead{Y-band SMF coupling($\%$)} & \colhead{I-band Strehl ($\%$)}}
	\startdata
	HD 89758 & M0 III  & 3.05  & 0.10 & 153/300 & 1-1.5 & 1.04/1.01 & 8/12 $\pm2$  &19/20  \\ 
	HD 97778 & M3 IIb & 4.63  & 1.05 & 300      & 1     & 1.04  & 12 $\pm2$    & 12    \\ 
	NSV 19434& M7     & 10.2  & 2.19 & 300      & -     & 1.01  & 11 $\pm2$    & 7     \\ 
	SA0 82686& M4     & 9.76  & 2.75 & 300/400  & -     & 1.02    & 22/23 $\pm2$ & 26/27 \\ 
	HD 113496& M3     & 7.50  & 2.75 & 400      & 0.8   & 1.09  & 22 $\pm2$    & 26    \\ 
	\enddata
	\tablecomments{Summary of observations including stellar parameters, telemetry, and relevant instrument measurements. Spectral type and apparent ($V$, $J$) magnitudes are quoted from SIMBAD. AO modes refers to the number of correcting elements applied on the adaptive secondary mirror. Missing entries in seeing data are due to corrupt data; however, during periods of stable seeing $\leq 1"$, up to 400 modes were used, as opposed to 153-300 modes when the seeing was in the range 1.0"-1.5" or less stable. Repeated observations with different AO modes are indicated with a slash. The AO loop frequency on all targets was 990Hz with 1x1 pixel binning. Strehl ratios and fiber coupling values are averaged over an entire observing set (5-30 min).} 
\end{deluxetable*}

\section{Procedure}
\label{sec:procedure}

The primary goal of this study is to: (i) report SMF coupling efficiency in the Y-band; and (ii) quantify specific fiber coupling losses. SMF coupling efficiency in the demonstrator system was measured by dividing the total power output through the SMF by the incident power measured at the SMF tip using a pair of calibrated photodiodes. This ratio provides an ``instantaneous" coupling efficiency, which serves a baseline measurement for understanding coupling losses and directly informs science performance predictions.

The theoretical SMF coupling efficiency, $\rho_{th}$, is computed from the normalized overlap integral between the telescope pupil function and the projected fiber mode in the pupil plane:
\begin{equation}
\label{eq:rho}
    \rho_{th} = \left|\frac{\braket{~E_{tel}(u)~|~E_{f}(u)~}}{||~E_{tel}(u)~||.||~E_{f}(u)~||}\right|^2,
\end{equation}
where $E_{\rm tel}(u)$ is the complex pupil function of the telescope and $E_{f}(u)$ the complex conjugated fiber mode propagated to the pupil plane~\citep{Ruilier2001}.

However, without full knowledge of the complex electric field, we must characterize fiber coupling losses experimentally using quantities that are measured in our system. To do this, we consider both dynamic and static sources of error. Dynamic terms include imperfect wavefront correction from the AO system (Strehl ratio) and residual tip-tilt errors. Static terms are characterized based on telescope pupil geometry, fiber misalignment, and non-common path aberrations (NCPA). We investigate these possible sources of loss using diagnostic images recorded contemporaneously with fiber coupling.

The effects of degraded throughput are considered using four separable terms derived from diagnostic data: 
\begin{equation}
\label{eq:rho_sep}
    \rho = \rho(Strehl)\rho(tip/tilt)\rho(tel)\rho(ncpa),
\end{equation}
where $Strehl$ is the Strehl ratio, $tip/tilt$ represents tip-tilt residuals, $tel$ is the telescope pupil geometry, and $ncpa$ represents static aberrations. Static errors include NCPA resulting from optical alignment errors, high order terms from optical surface quality, as well as defocus errors ($Z4$) introduced by fiber misalignment. Each parameter in Equation~\ref{eq:rho_sep} is quantified as an equivalent fiber coupling value, e.g. measured tip-tilt residuals in the focal plane are translated to a fiber coupling coefficient of $\rho_{\rm tip/tilt}=0.84$. These parameters derived from diagnostic data are used to inform the complex electric field model of the fiber and telescope beam at the pupil plane. 

Finally, fiber coupling results from this analysis are used to inform simulations of Doppler spectroscopy under similar observing conditions. As fiber coupling losses reduce observational efficiency and degrade RV precision, characteristic values are included in end-to-end instrument simulations that model signal to noise ratio of a simulated stellar spectrum. These simulations calculate the photon-noise-limited RV precision as a function of apparent magnitude for a diffraction-limited spectrograph at the LBT. These predictions are used to derive the relationship between fiber coupling efficiency and limiting apparent magnitude.  

\section{Observational Summary}
\label{sec:summary}

We report on-sky measurements from April 2016 for five different stars observed with apparent magnitudes ranging from $V=(3.05-10.2)$ in seeing conditions from $\theta=(0.8-1.5)$". Observations of bright, mid-late M-type stars were recorded to provide high signal-to-noise (SNR) ratio in the fiber coupling channel, $\lambda =~(970-1065)$~nm, while varying the SNR in the wavefront sensing channel, $\lambda = (400-970)$~nm. Data sets for each star include simultaneous measurements of the Y-band SMF coupling efficiency, high speed imaging data in the $I$-band, and wavefront sensor (WFS) telemetry. 

A summary of results from the final night of the 2016 observing run is shown in Table~\ref{tab:obs_sum}. Observations of each star were limited to 2-30 minutes to minimize the effects of instrument flexure and changing airmass as there was no active beam stabilization or continuous atmospheric dispersion correction in the prototype system. Measured Y-band SMF coupling efficiency ranged from 8~-~23\% with I-band Strehl ratios from 7-20\% while using 153-400 AO modes (400 was the maximum number of modes in 2016). Although the apparent visual magnitude differed by as much as $\Delta V = 7.15$ for the various targets, AO correction was limited primarily by atmospheric seeing conditions and not flux on the WFS. 

To achieve AO correction, the LBT adaptive secondary mirror has N=672 actuators and operated with a bandwidth of 990Hz in 2016~\citep{Esposito2010}. Recently at the LBT, the SOUL AO upgrade has increased the pupil sampling in the WFS and improved the bandwidth~\citep{Pinna2016}. For comparison, other AO systems with plans to use SMF units include the Keck Planet Imager and Characterizer (KPIC), which has N=1156 actuators (34x34) and 1500Hz bandwidth, as well as the PALM3000 at the Hale Telescope which has N=2500 actuators and 2000Hz bandwidth~\citep{Mawet2017,Dekany2013}. Future extremely large telescopes will have significantly more actuators to account for larger primary mirrors: the TMT is will include approximately 8000 actuators, whereas both the ELT and GMT have plans to use 5000 actuators respectively~\citep{Hippler2018}.

\section{Fiber Coupling Efficiency}
\label{sec:SMFcoupling}

\begin{figure*}
    \centering
    \includegraphics[width = 7in]{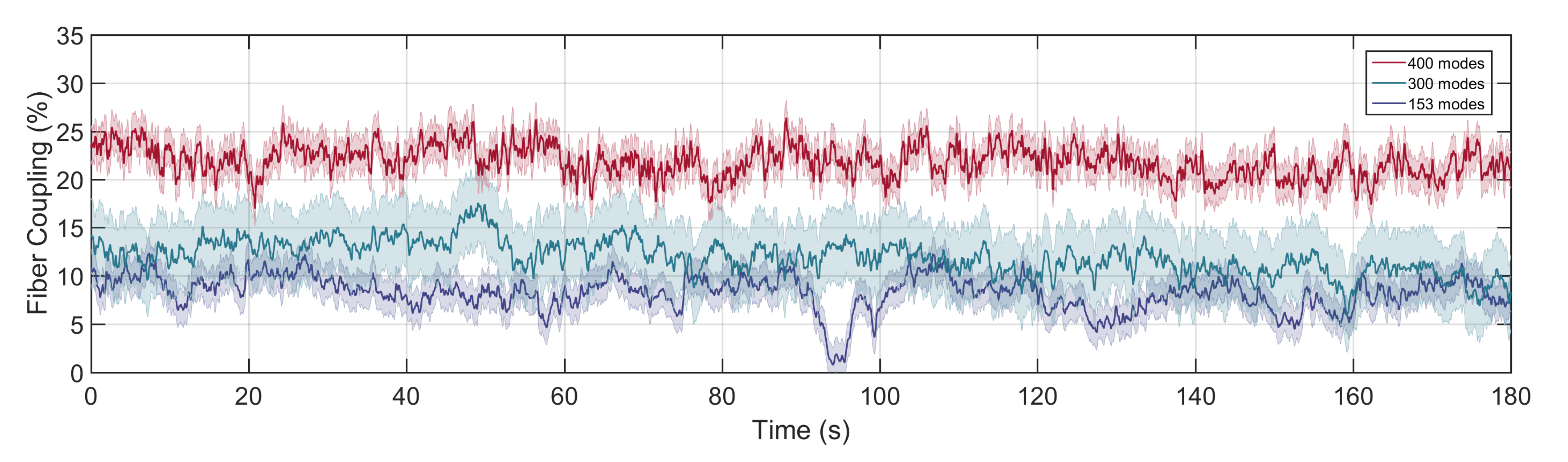}
    \caption{Instantaneous coupling efficiency of three observations: HD~113496 (\textcolor{dred}{\textit{Red}}) with 400 AO modes correction, HD~89758 with 300 AO modes (\textcolor{dgreen}{\textit{Green}}), and HD~89758 with 153 AO modes (\textcolor{dpurp}{\textit{Purple}}). Shaded regions indicate uncertainty in the instantaneous measurement.}
    \label{fig:rho}
\end{figure*}

The instantaneous fiber coupling efficiency of three observations are shown in Figure~\ref{fig:rho}. In each case, the peak coupling (25\%, 17\% 12\%) occurs near the start of the observation and degrades with time. This is expected due to the lack of active beam stabilization to correct for beam wander and degrading atmospheric dispersion correction. The average coupling efficiencies over 180 seconds for the 400 modes, 300 modes, and 153 modes are 23\%, 12\% and 8\% respectively. As one would expect, the average fiber coupling efficiency is related to the number of AO modes (Figure~\ref{fig:rho} and Table~\ref{tab:obs_sum}). This follows our understanding of theoretical SMF injection as higher order correction on the adaptive secondary mirror should provide better wavefront correction, improving the Strehl ratio and fiber coupling efficiency.

Although more correcting modes with HD~113496 improves the average coupling efficiency, significant temporal variability is present as both stars show similar levels of dynamical loss. The relative scatter is similar in each case with $\sigma_{\rm rms} = 1.57\%$ at 400 modes, 1.73\% at 300 modes and 1.75\% at 153 modes over the 3 minute interval. This suggests the cause of dynamical fiber coupling loss may be dominated by effects independent of AO correction. However, the effects of high order wave front correction and tip-tilt residuals are difficult to disentangle from each other in fiber coupling efficiency data. Therefore we require additional diagnostic data to break the degeneracy and quantify the impact of wave front correction and tip-tilt individually. 

\section{Pupil Geometry}
\label{sec:Pupil}
A typical telescope pupil is modeled as a large circular aperture from the primary mirror with a smaller central obstruction from the secondary mirror. Support structures for the secondary and tertiary mirrors, sometimes referred to as spiders, are also visible in the pupil. To first order, this pupil geometry causes diffraction only. Diffraction displaces energy from the PSF core into the Airy rings where the amount of displaced energy increases with the ratio of the secondary to primary mirror diameter. These effects reduce mode-matching, placing an upper limit on SMF coupling efficiency. 


In the case of a perfectly corrected wave front, diffraction effects from the secondary limit the maximum achievable fiber coupling efficiency to $\rho_{\rm tel}~=~0.78$. This theoretical value of $\rho$ represents the maximum achievable SMF coupling and is used as a normalization factor throughout the remainder of this work. We calculate $\rho$ using Equation~\ref{eq:rho} with a secondary mirror blocking fraction of $11\%$ and fiber mode-field diameter (MFD) of $5.8 \; \mu m$. Additional diffraction caused by support spiders removes light from the PSF core.  Both the secondary and tertiary support structures at the LBT have a (projected) width of approximately 8\% that of the secondary diameter and form an 18-degree wedge angle. Supports degrade the maximum fiber coupling efficiency by 2-3\%.

Light located in the Airy rings can’t be coupled into SMFs with standard optics, accounting for $\approx$ 20\% of the total power. Phase Induced Amplitude Apodization (PIAA) optics remap starlight from the Airy rings back into the PSF core and have been tested with SMF injection~\citep{Jovanovic2017}. PIAA optics are considered a future upgrade for SMF injection at the LBT in the case that fiber coupling is limited by diffraction as opposed to AO correction. Although PIAA optics combined with a flat wavefront improve coupling,  Strehl ratios above  80\% are warranted~\citep{Sivaramakrishnan2001}.

\section{Strehl ratio}
\label{sec:Strehl}
The Strehl ratio is an extremely useful parameter when assessing SMF injection as it is proportional to the maximum achievable coupling efficiency \citep{Wagner1982,Ruilier2001}. The Strehl is also a convenient parameter to summarize wavefront errors (WFE) and can be computed from contemporaneously recorded imaging data. Importantly, we distinguish between the maximum achievable coupling efficiency and measured coupling as other instrumental and telescope effects may degrade fiber coupling without impacting the Strehl ratio. 

In the demonstrator system, the Strehl ratio is measured at $\lambda = 800$nm by computing a theoretical diffraction pattern using a simulated LBT pupil function according to our PSF sampling ($f \lambda /D$ = 6.9 pixels). Both measured and simulated frames are tip-tilt centered, normalized and fit with a 2D Gaussian function to find the maximum amplitude. The PSF centroid is found using the flux weighted image center, followed by Fourier shift to center the PSF in the image frame to remove tip-tilt. This is followed by a 2D Gaussian fit to the PSF core in each frame to calculate the PSF amplitude. The Strehl ratio is then calculated as:
\begin{equation}
\label{eq:SR}
    SR = \frac{I(x=0)}{\Sigma I}\frac{\Sigma P}{P(x=0)},
\end{equation}
where $I$ is the measured intensity and $P$ is the simulated intensity of the diffraction pattern and $x=0$ is the center of the image \citep{RobertsJr.2004}. 

\subsection{Instantaneous Strehl ratio}
\begin{figure*}
    \centering
    \includegraphics[width =7in]{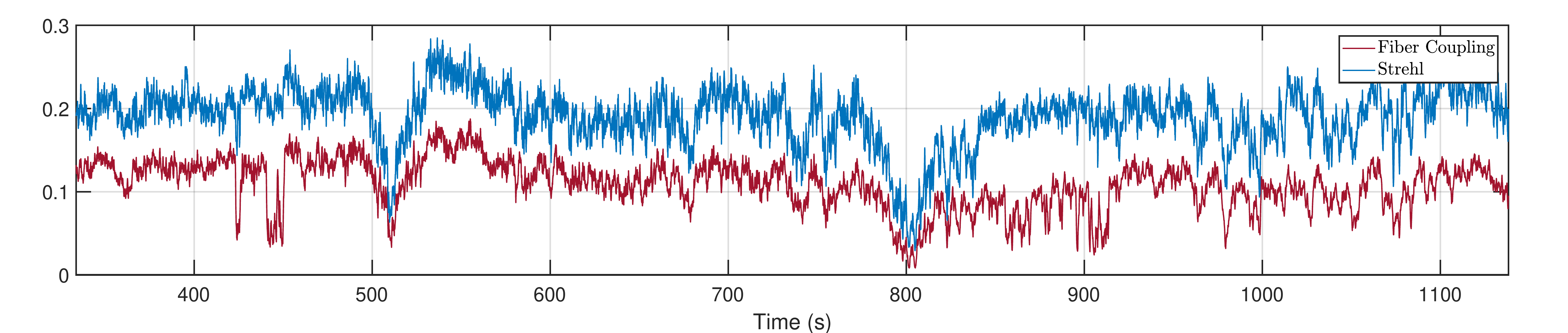}
    \caption{I-band Strehl Ratio an fiber coupling time series over a 13 minute observation for HD 89758, Set V (see Table~\ref{tab:deets} for full details). The correlation coefficient is computed to be $\gamma_{\rm Strehl}=0.77$, indicating a strong positive correlation between instantaneous fiber coupling efficiency and Strehl ratio.}
    \label{fig:corr}
\end{figure*}

A time series of Strehl ratio values and fiber coupling for HD~89758 at 300 modes is shown in Figure~\ref{fig:corr}. The Strehl ratio (blue) is overlaid with normalized fiber coupling (red) after being temporally synchronized. The Strehl ratio average is 18\% with an rms of 5\%, with distinct features at 500 and 800 seconds. In a similar fashion to the instantaneous fiber coupling in Figure~\ref{fig:rho}, Strehl ratio variations are a combination of low amplitude jitter and higher amplitude drifts.

To quantify the linear dependence of temporal fluctuations in Strehl ratio and fiber coupling data we calculate the correlation coefficient, $\gamma_{\rm Strehl}$. A coefficient of 0 indicates no correlation while $\pm1$ indicates completely positive correlation/anti-correlation. To compute $\gamma_{\rm Strehl}$, the Strehl ratio and fiber coupling efficiency are normalized by mean subtracting and dividing by the standard deviation. Next, data are time synchronized using cross-correlation to remove lag-time offset introduced by slight clock time differences in each data set. A moving average filter is also applied to match the Strehl ratio sampling to fiber coupling bandwidth. Visually the signals in Figure~\ref{fig:corr} show a very clear positive correlation, especially considering the slow drift of each signal over the full 13 minutes. Formally, a correlation coefficient, $\gamma_{\rm Strehl} = 0.77$, is calculated for HD 89759, data set V (see Appendix~\ref{sec:deets}) where $\gamma\geq 0.7$ is typically considered very strong statistical dependence.  

The correlation between Strehl ratio and fiber coupling is calculated for all observations in Table~\ref{tab:deets}. All stars indicate fiber coupling efficiency is strongly correlated with instantaneous Strehl ratio, with an ensemble average $\gamma_{\rm Strehl}\geq0.70$. This is in agreement with the theoretical relationship between fiber coupling and Strehl and confirms the `instantaneous' Strehl ratio accounts for the majority of low frequency temporal variations in the fiber coupling efficiency for all observed stars with varying number of AO modes, seeing, airmass, and brightness. 

\subsection{Limitations set by Strehl ratio}
Comparing Strehl ratios with fiber coupling provides an estimate of how closely each observation was to the theoretical limit set by AO correction. To more easily quantify performance, we rescale the Strehl from the $I$-band to the fiber coupling wavelength band ($Y$-band) and normalize the fiber coupling efficiency to the maximum possible efficiency according to the telescope pupil geometry in Equation~\ref{eq:rho} (i.e., a maximum of 78\%).\footnote{The measured I-band and rescaled Y-band Strehl ratios are in agreement with Strehl ratios at the LBT presented in \citet{Pinna2016}}. In doing so, we establish a 1:1 relationship between the adjusted Strehl ratios and fiber coupling.  

\begin{figure}
    \centering
    \includegraphics[width = 3.3in]{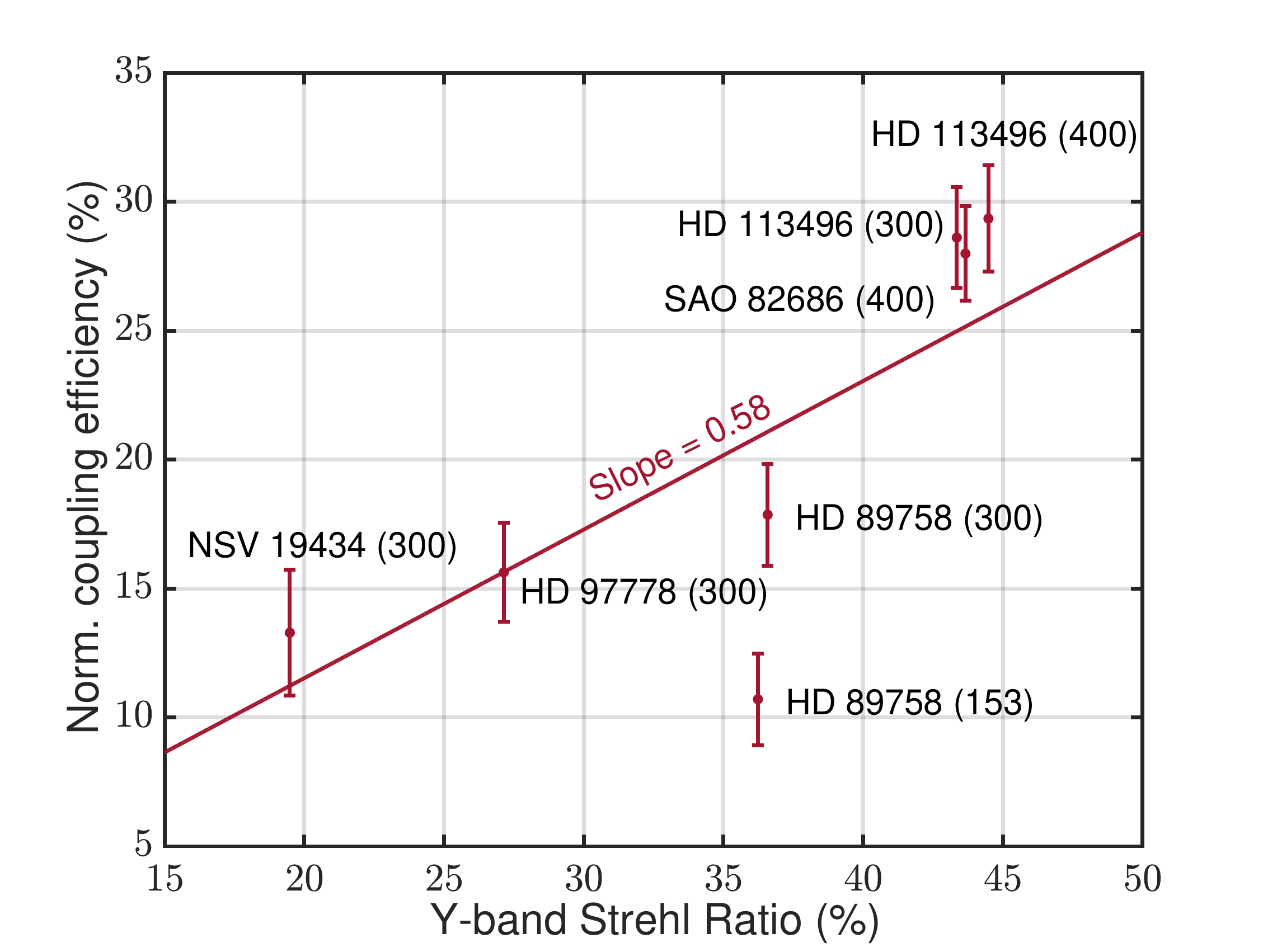}
    \caption{Average Y-band coupling efficiency vs. Y-band Strehl ratio for each observation in Table~\ref{tab:obs_sum}. Fiber coupling efficiency is normalized to the maximum theoretical efficiency (78\%) and the Strehl ratio has been re-scaled to estimate the Y-band Strehl ratio from I-band measurements. A linear fit indicates the instrument coupled approximately 60\% of the available light.}
    \label{fig:SR_fit}
\end{figure}

Results are shown in Figure~\ref{fig:SR_fit}. In general, observations with 400 AO modes performed more closely to the theoretical limit than 300. For well corrected  stars, the Strehl was $\sim$45\% which corresponds to a non-normalized maximum fiber coupling efficiency of $\rho_{\rm max}=0.35$ with the pupil geometry of the LBT. Observations with worse AO correction had a Strehl ranging from 20-36\% with $\rho_{\rm max}=0.15-28$. Interestingly we find HD~89758  coupled significantly less light using 153 AO modes than with 300 modes despite achieving similar Strehl r.atios. Following the formalism of Equation~\ref{eq:rho_sep}, we calculate $\rho_{\rm Strehl} = 0.20-0.45$.

Fiber coupling degradation beyond values predicted by the Strehl can be estimated by computing a linear fit, as shown in Figure~\ref{fig:SR_fit}. A fit value of $58~\pm~8\%$, indicates the demonstrator typically coupled just over half of the available light as predicted by the Strehl ratio. This suggests additional factors degraded the fiber coupling efficiency by an additional 40\%. Furthermore, while the measured correlation is very high, additional effects impacting fiber coupling are evident in Figure~\ref{fig:corr}, where the fiber coupling deviates significantly from the Strehl ratio at 440 seconds and 850-920 seconds. These unaccounted variations are most likely the result of large amplitude misalignment of the PSF and fiber and are investigated in \S\ref{sec:TT} and \S\ref{sec:static}. 


\section{Tip-tilt residuals}
\label{sec:TT}
Residual tip-tip errors may be the consequence of uncorrected atmospheric turbulence, over-driving the AO system, telescope vibrations, or combinations thereof. To understand the effects of tip-tilt motion we investigate the temporal and spatial domains of the PSF. The PSF centroid is found using a 2D Gaussian fit to the PSF core in each frame. We characterize the tip-tilt signal in terms of a scatter amplitude, $\sigma_{\rm tip/tilt}$, and drift term, $\delta$, where $\delta$ is measured by applying a low-order polynomial fit. The slope is a measure of the angular centroid `drift per unit time' and $\sigma_{\rm tip/tilt}$, is calculated by taking the rms of centroids after subtracting the mean.

\begin{figure*}
    \centering
    \includegraphics[width =7in]{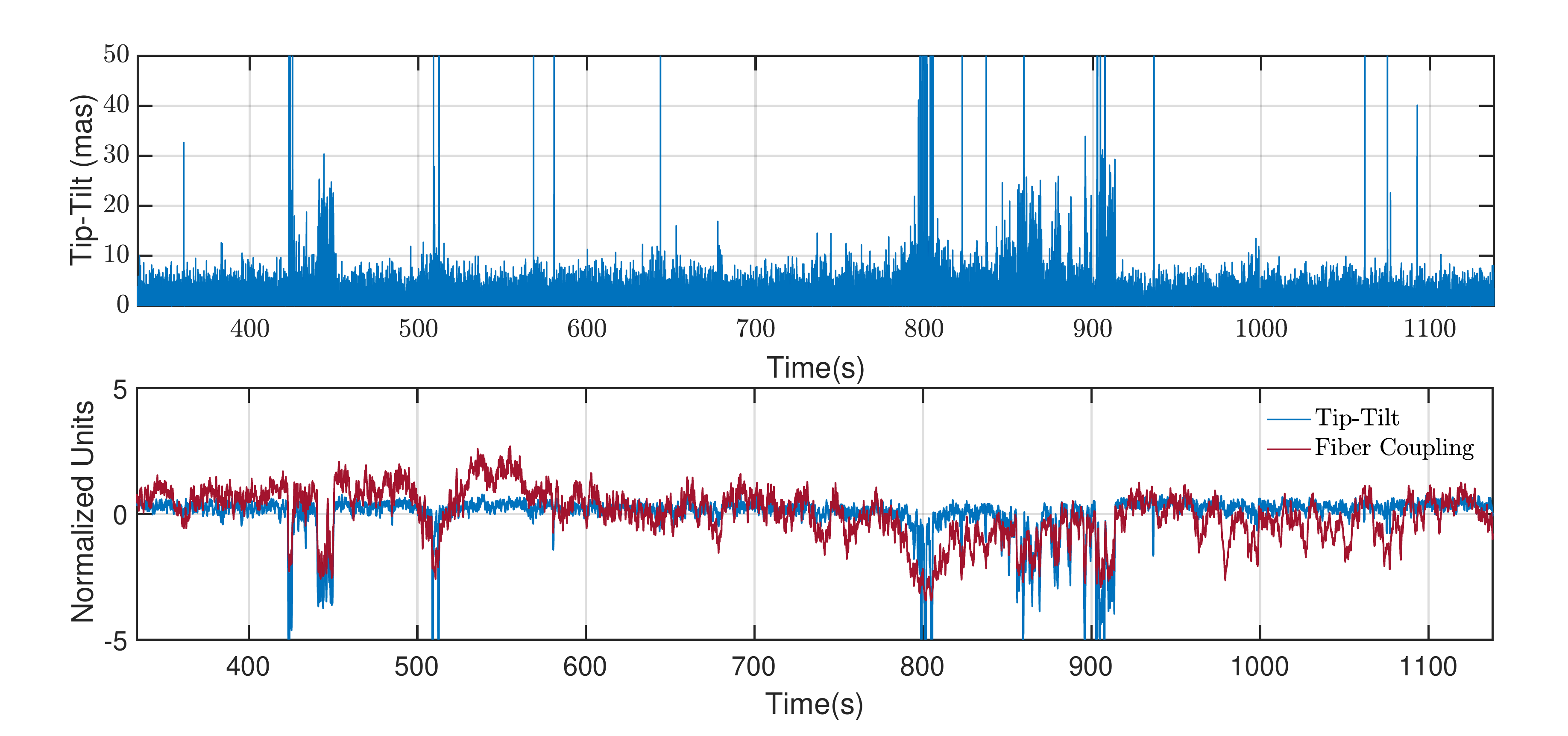}
    \caption{\textbf{Top}: Tip-tilt time series calculated from HD~89758~SetV PSF centroids. Tip-tilt peaks located at 400, 800, and 900 seconds are due to poor AO correction.  \textbf{Bottom}: Normalized Tip-Tilt and Fiber coupling signal. Each signal is mean subtracted, normalized by the standard deviation, and cross correlated to remove sensor lag offset before computing the correlation coefficient. A correlation coefficient of $\gamma_{\rm tip/tilt}=0.56$ suggests a moderate linear relationship between fiber coupling and PSF location.}
    \label{fig:tt}
\end{figure*}

The tip-tilt time series for HD89758~SetV is shown in the top panel of Figure~\ref{fig:tt} with a sampling frequency of $f=120$ Hz. HD89758~SetV is measured to have a drift of $\delta= 0.1$ mas/minute and scatter amplitude of $\sigma_{\rm tip/tilt}= 10.5$ mas measured over 13 minutes. In contrast to the Strehl ratio, the tip-tilt signal does not contain significant low frequency components ($f\leq 1$ Hz).

Day-time engineering tests were analyzed to understand the LBT AO system and telescope vibrations at higher SNR, and sampling frequency without the atmospheric turbulence effects. By closing the AO loop using a bright $\lambda=1064$ nm calibration source, we sample the PSF at $f=3$ kHz. Day-time tests are referred to as closed-dome SX and DX (Appendix~\ref{sec:deets} and correspond to the left and right side of the LBT respectively. 

A summary of centroid drifts is shown in Table~\ref{tab:deets}. The average drift is $\delta = 0.74\pm 0.88$ mas/minute for on-sky data and $\delta = 2.43$ mas/minute for closed dome measurements. For both on-sky and closed dome pixel scales (2.87 and 3.92mas/pixel respectively), $\delta \leq 2-3$ mas/minute corresponds to an extremely small motion at the sub-pixel level. Thus the average PSF centroid remained in an almost constant location throughout short observations, and atmospheric turbulence and telescope elevation (i.e. instrument flexure) did not cause significant PSF drift over fiber coupling timescales. 

The correlation between fiber coupling and tip-tilt signal for HD~89758~SetV is shown in the bottom panel of Figure~\ref{fig:tt} where the centroid sampling has been down-sampled to match fiber coupling measurements. The tip/tilt correlation in Figure~\ref{fig:tt} is measured to be $\gamma_{\rm tip/tilt}=0.56$. An ensemble average of $\gamma_{\rm tip/tilt}=0.46$ for all observations which suggests a positive, moderate overall correlation between fiber coupling and PSF location.  Occasionally, the tip-tilt signal also accounts for some distinct high amplitude fiber coupling losses, not captured by the Strehl ratio. Examples of these regions are shown in the bottom panel of Figure~\ref{fig:tt} at 450 seconds and 850-920 seconds where dominant tip-tilt losses are captured by the fiber coupling data. 

Overall, we find the measured tip-tilt residuals do not correlate as strongly with fiber coupling as the Strehl ratio. However, high frequency components present in the tip-tilt signal, but not present in fiber coupling data, are likely due to a bandwidth limitation of the photodiode electronics ($f\leq 1-5$ Hz). Further observations with higher sampling rates of fiber coupling data are expected to provide a better understanding of high frequency ($f\geq 1$Hz) tip-tilt variations. With a higher fiber coupling bandwidth (i.e. $f\geq 30$ Hz) we would expect to capture more high frequency information through the fiber and perhaps measure a stronger correlation between PSF position and fiber coupling.  

\subsection{Limitations set by Tip-tilt}

\begin{figure}
    \centering
    \includegraphics[width =3.3in]{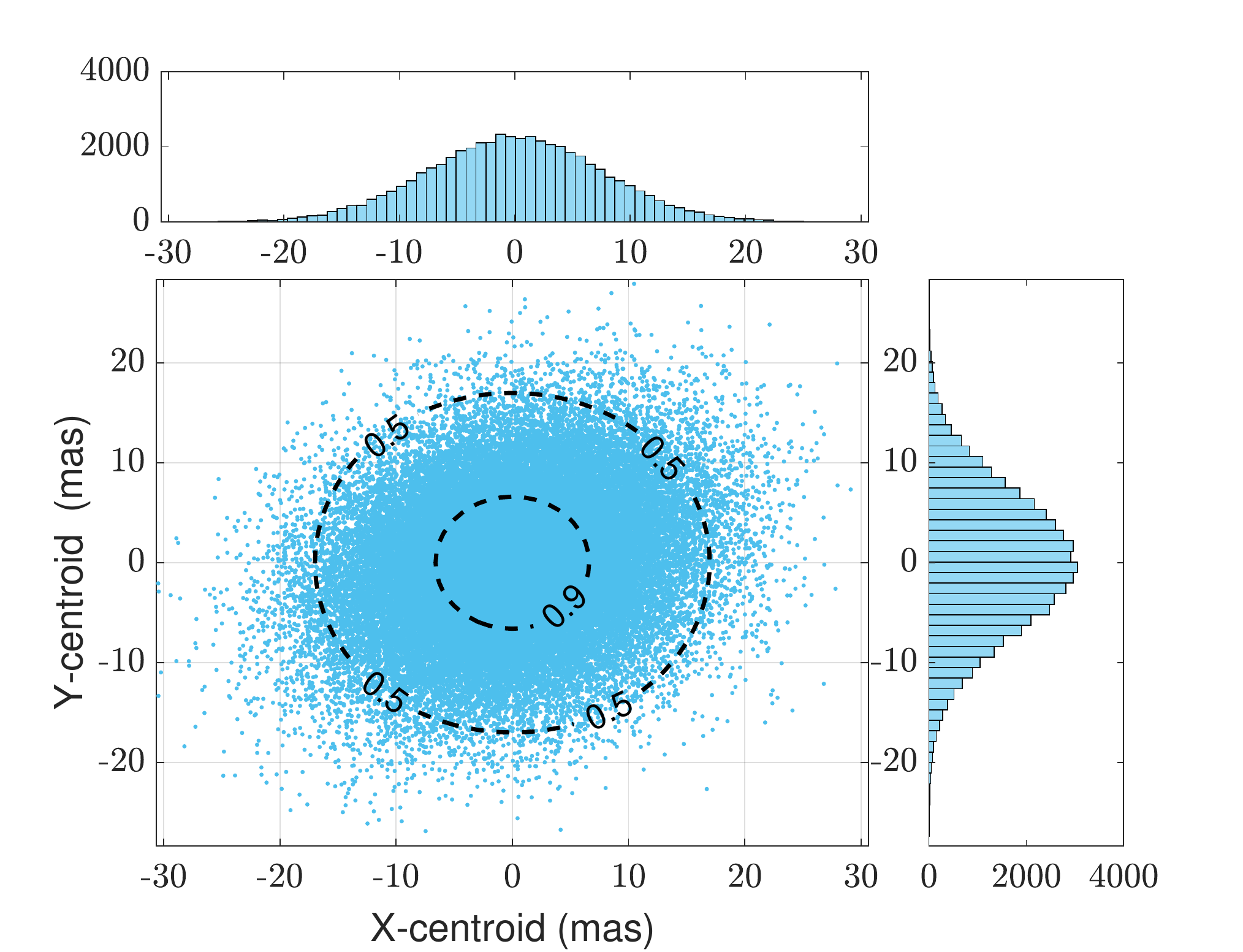}
    \caption{Centroid scatter plot for SA0 82686 II~Set. $\sigma_x = 7.66$~mas and $\sigma_y = 7.09$~mas correspond to the standard deviation of the centroid scatter. The maximum deviation in x, y directions is 59.61, 57.76 mas respectively. Horizontal and vertical histograms indicate the spatial distribution is near Gaussian.  Contour lines (--) show normalized coupling efficiency of 0.90 and 0.50 for a given tip-tilt amplitude.}
    \label{fig:scatter}
\end{figure}

To visualize tip-tilt amplitudes we show the 2D centroid scatter for SAO 82686~SetII in Figure~\ref{fig:scatter}. The centroid distribution is approximately a circular Gaussian with a slight elongation at 45$^\circ$. An average characteristic scatter corresponding to the $1\sigma$ Gaussian width parameter is $\sigma_{\rm tip/tilt} = 7.37$ mas with the $\mbox{FWHM} = 17.4$ mas. For comparison, the angular projection of the SMF mode-diameter is about 42 mas at $\lambda=1\mu$m and $\sigma_{\rm tip/tilt} \approx 1/6$ fiber diameters.

For a fiber that is perfectly centered on the mean scatter position, we compute the normalized fiber coupling coefficient, $\rho_{\rm tip/tilt}$, in the presence of tip-tilt errors. To do this we first compute an equivalent WFE ($Z2,Z3$) in the pupil plane that results in the measured $\sigma_{\rm tip/tilt}$ in the focal plane, and then calculate $\rho$ according to Equation~\ref{eq:rho}. Using this technique we calculate contour lines of 1 and 2.33$\sigma_{\rm tip/tilt}$ which correspond to $\rho_{\rm tip/tilt}=$ 0.90 and 0.50 respectively and are shown in Figure~\ref{fig:scatter} . 

A summary of centroid scatter measurements can be found in Table~\ref{tab:deets}. The characteristic on-sky scatter has an ensemble average of $\sigma_{\rm tip/tilt} = 8.8 \pm 2.3 $mas which is equivalent to 1/5th of a fiber diameter and a fiber coupling loss of $16 \pm 7\%$. Following the formalism of Equation~\ref{eq:rho_sep}, we conclude $\rho_{\rm tip/tilt} = 0.77-0.91$.

\subsection{Telescope Vibrations}

\begin{figure*}
    \centering
    \includegraphics[width =6.5in]{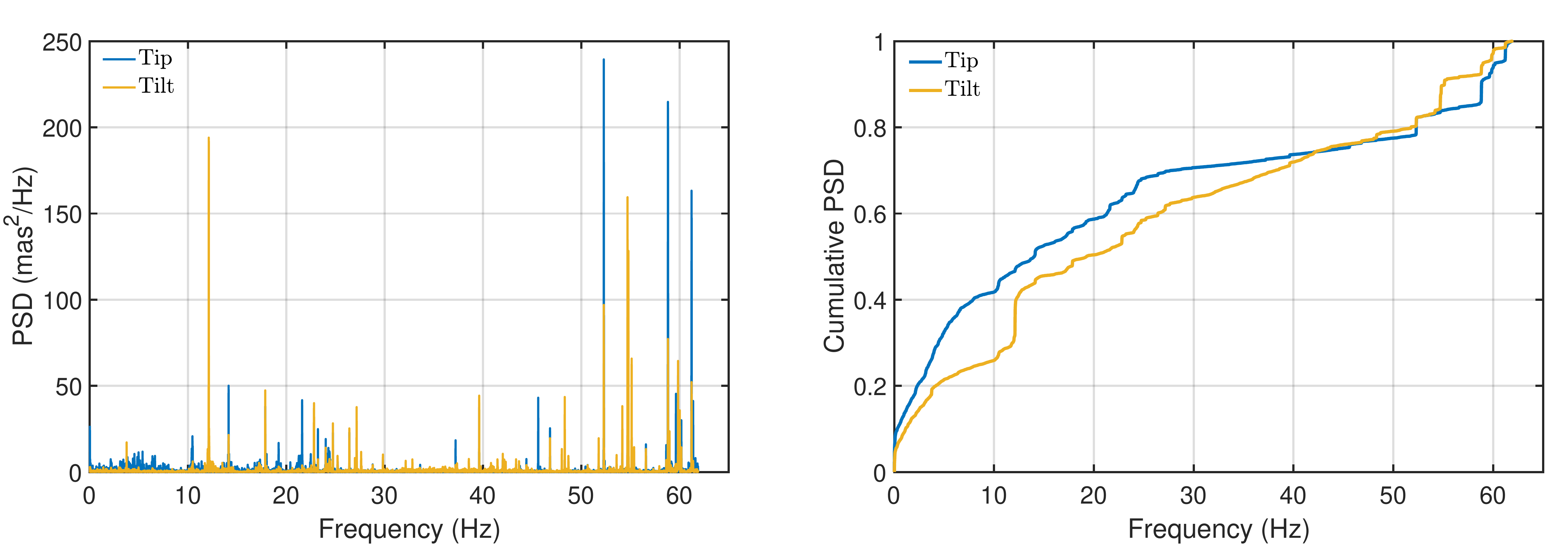}
    \caption{\textbf{Left}: Tip-tilt power spectral density of HD89758 with 300 AO modes. Sharp peaks are present in both axes from 10-30 Hz and 50-60 Hz. \textbf{Right}: Normalized cumulative tip-tilt PSD. Approximately 70\% of the energy is contained within the 0-30 Hz band. The remaining energy is concentrated from 50-60 Hz.}
    \label{fig:vibrations}
\end{figure*}

Vibrations at the LBT are present on the telescope structure supporting the primary, secondary and tertiary mirrors and are not entirely corrected by the LBT AO system. A power spectrum computed from our on-sky x- and y-centroid time series is shown in the left panel of Figure~\ref{fig:vibrations}. We identify sharp vibration signatures in the 10-30Hz band as well as the 50-60Hz band. This strongly indicates residual tip-tilt errors in the demonstrator system were impacted by telescope vibrations and supports findings suggested by the closed-dome and on-sky tip-tilt amplitudes. 

The cumulative PSD (right side of Figure~\ref{fig:vibrations}) reveals 60-70\% of the vibration energy is contained in the 0-30 Hz band and an additional 30\% in the 50-70 Hz band. This distribution of energy calculated from centroid positions in demonstrator focal plane is in agreement with both LBT WFS and accelerometer measurements in \citet{Escarate2018}. This further validates our findings that tip-tilt residuals originate from telescope vibrations. 

To address telescope vibrations, mitigation techniques within the fiber injection module will be tested in 2019 with a goal of targeting vibrations in the 0-30 Hz band. These test runs will include the use of a fast tip-tilt mirror and IR quadcell photodiode. Beam stabilization with a quadcell will also be compared to tip-tilt correction using a prototype photonic device described in \citet{Hottinger2018}.   

\section{Static Aberrations}
\label{sec:static}
Static aberrations are typically introduced by non-common optical paths between the wavefront sensor (WFS) and science instrument. In this instance however, we refer to NCPA as the difference in WFE between the demonstrator imaging channel and fiber channel. The slow lenses in the imaging channel used to compute the Strehl impart a small WFE and are easier to align than the highly curved surfaces of the fast lenses in the fiber channel. Additional static aberrations may also be introduced by mismatches between the fiber tip and incident beam. Thus, WFE in the imaging channel should be considered a lower limit on the WFE at the fiber tip.  

\begin{figure}
    \centering
    \includegraphics[width =3.2in]{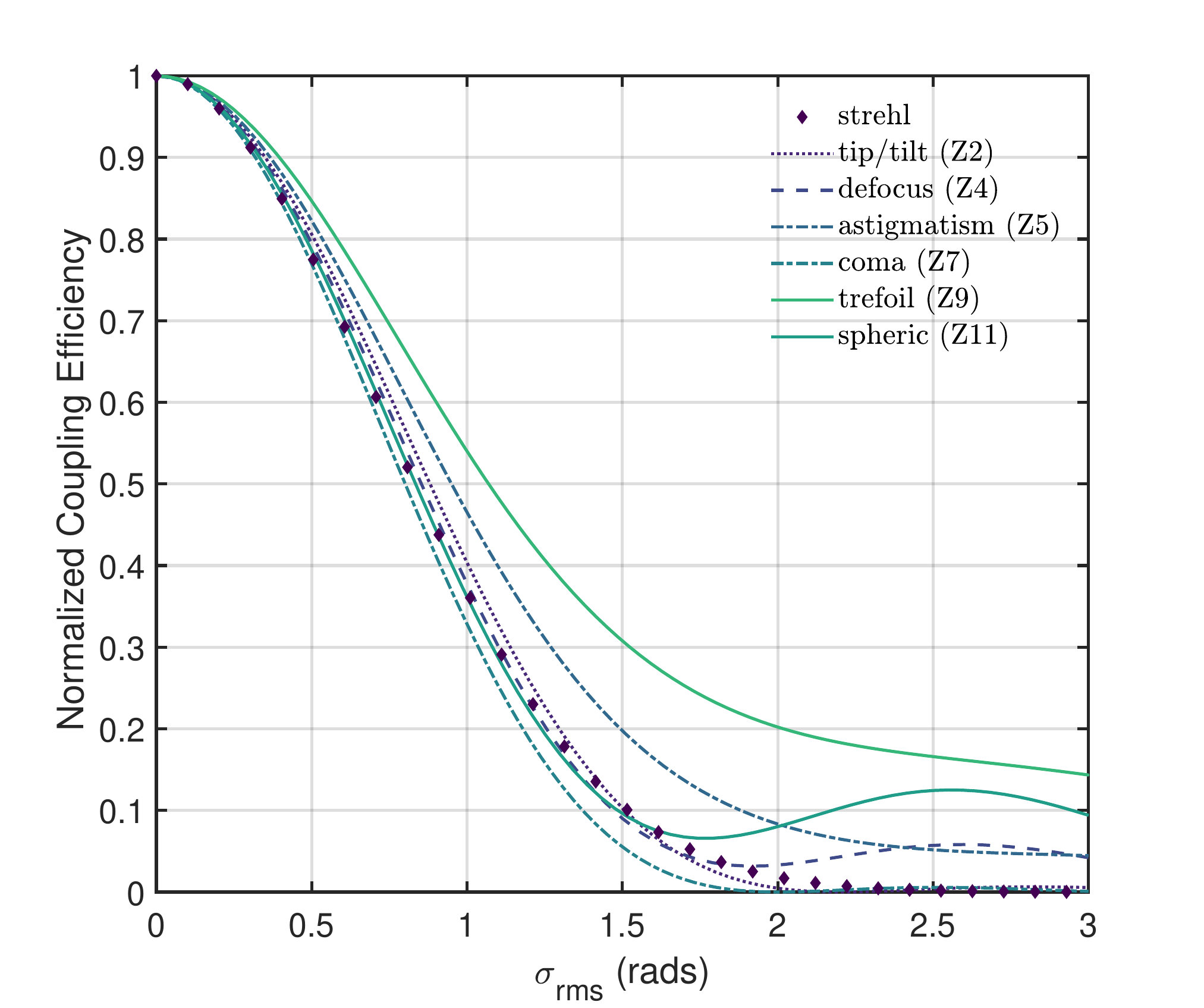}
    \caption{Fiber coupling vs. rms WFE ($\sigma_{rms}$) as a function of pure Zernike modes and the Strehl ratio. Zernike modes are numbered according to the Noll index \citep{Noll1976}.} 
    \label{fig:static}
\end{figure}

We approximate the sensitivity of the system to static aberrations by calculating theoretical coupling efficiency for the first several Zernike modes. To do this, we propagate the SMF Gaussian mode from the focal plane to the pupil plane and calculate the overlap integral of the fiber mode with the telescope pupil function in Equation~\ref{eq:rho}. Using this methodology, physical fiber alignment errors in the focal plane such as lateral and focus misalignment manifest as low order static aberrations. Figure~\ref{fig:static} shows the relationship between rms WFE, $\sigma_{\rm rms}$, and normalized fiber coupling efficiency. Most Zernike terms degrade fiber coupling similarly whereas Astigmatism (Z5) and Trefoil (Z9) degrade more slowly with WFE. The Strehl ratio is calculated using the Mar\'{e}chal approximation, where $\sigma_{\rm rms}= 0.5, 1$ is equivalent to $sr = \rho_{\rm ncpa} =$ ($0.80$, $0.40$) \citep{Ross2009}. 

The remaining fiber coupling loss not accounted for by the Strehl or tip/tilt residuals can be explained with either a lateral or focus misalignment of the SMF. For an exclusively lateral or focus error to induce a 20\% injection loss, static WFE with an amplitude of $\sigma_{\rm rms}=0.5$ in either Z(2,3) or Z4 could be present. In the case of a pure lateral misalignment, this WFE translates to $\Delta{x}=1.4 \; \mu$m in the focal plane, or about 1/4th of the physical fiber core diameter. The equivalent defocus error in waves is calculated according to:
\begin{equation}
    D = \frac{\Delta z}{8\lambda F^2},
    \label{eq:waves}
\end{equation}
where $\Delta z$ is the physical distance moved along the optical axis, $\lambda$ is the wavelength, and F is the focal ratio. Using Equation~\ref{eq:waves}, and scaling $\sigma_{\rm rms}=0.5$ to P-V WFE in waves, a defocus error of $\Delta z=125 \; \mu$m degrades fiber coupling efficiency by 20\%. The total static WFE however, is likely a combination of focus and alignment errors, combined with higher order surface errors. Following the formalism of Equation~\ref{eq:rho_sep}, we estimate $\rho_{\rm ncpa} \approx 0.8$. 

\subsection{Phase Retrieval}
\label{sec:MGS}

\begin{figure*}
    \centering
    \includegraphics[width =6.5in]{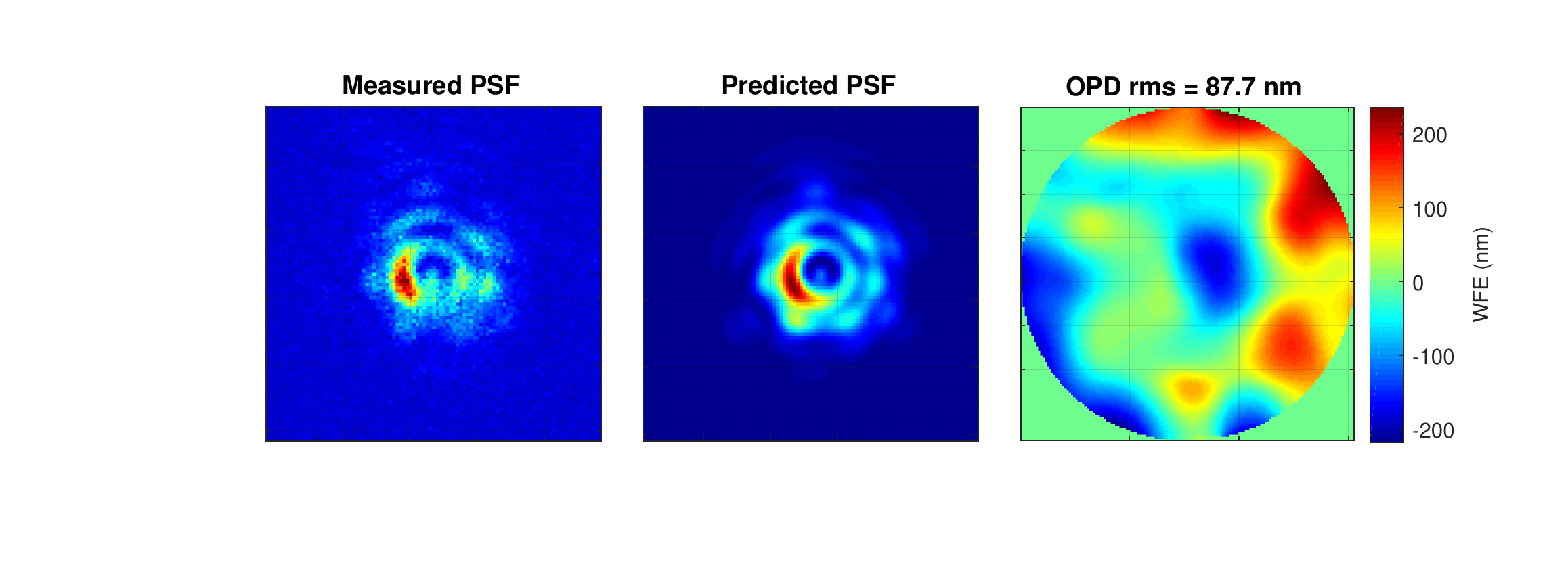}
    \caption{\textbf{Left}: Measured PSF at the LBT with 1 wave (12mm) of defocus using a $\lambda=1064$ nm calibration source. \textbf{Middle}: Predicted PSF from the MGS algorithm \textbf{Right}: Reconstructed WFE in the pupil plane.}
    \label{fig:MGS}
\end{figure*}

NCPA's degrade Strehl ratio and reduce fiber coupling efficiency. To overcome this problem, phase retrieval algorithms can be used to reconstruct the electric field phase after the WFS and apply appropriate offsets to the adaptive secondary mirrors for optimal correction in the science instrument. In the fiber injection system, these algorithms can be implemented in the fiber channel and imaging channel to correct for static aberrations and boost coupling efficiency closer to the theoretical maximum.

The Modified Gerchberg-Saxton (MGS) phase retrieval algorithm is a method that iteratively reconstructs the electric field phase by introducing known perturbations (diversity) to the system and imposing constraints on measured intensity~\citep{Green2003}. The algorithm provides a numerical estimate of the optical path difference (OPD). MGS has been demonstrated to significantly boost the Strehl ratio in astronomical applications including at Palomar Observatory\citep{Burruss2010}. 

We demonstrate the principle of phase retrieval at the LBT using a calibration source and camera that scans through a focal plane located after the WFS. Symmetric pairs of images are recorded around focus to introduce a known diversity and constrain the electric field amplitude. The OPD is reconstructed by iteratively propagating between the out of focus plane and in-focus plane for each image. After each propagation, an estimate of the phase is obtained as the known amplitude at each plane is re-imposed \citep{Bikkannavar2010}. Repeating this process allows the MGS algorithm to converge on an estimate of the OPD. 

Figure~\ref{fig:MGS} shows the comparison between a measured and predicted out of focus PSF. The right panel shows the OPD with $\sigma_{rms} = 87.7$nm, which is equivalent to a Strehl ratio of 0.76. To validate this method, we compare to the in-focus PSF Strehl ratio which is measured to be 0.83. Future experiments will aim to improve phase retrieval measurements by increasing diversity and applying offsets to the adaptive secondary mirror to correct the measured OPD. This approach will facilitate maintaining optimal image quality over seasonal thermal changes and  realignments of the incident beam to the fiber injection system. 

\section{Radial Velocity impact}
\label{sec:RV}
We investigate the impact of fiber injection losses on RV precision for iLocater at the LBT using the results from previous sections. RV uncertainties are calculated from end-to-end instrument simulations of the stellar spectrum across iLocater's 36 spectral orders \citep{Bechter2018,Bechter2019}. The simulations utilize fiber coupling coefficients: $\rho_{tel}$, $\rho_{ncpa}$, $\rho_{tip/tilt}$ as well as the modeled Strehl ratio to determine a variable chromatic coupling efficiency for a given spectral type and apparent magnitude. The variable coupling efficiency is combined with the static iLocater instrument throughput terms to determine the signal to noise ratio (SNR) of a given spectrum.

The photon-noise-limited RV precision, $\sigma_{\rm ph}$, is estimated using the formalism of \citet{Butler1996},
\begin{equation}
\label{eq:phnoise}
\sigma_{\rm ph}=\frac{1}{\sqrt{\sum \left(\frac{dI/dV}{\epsilon_I} \right)^2}},
\end{equation}
where $dI/dV$ is the slope of the measured stellar intensity as a function of wavelength (expressed in velocity units) and $\epsilon_I=\sqrt{N_{\rm ph}}/N_{\rm ph}$ is the fractional Poisson error at pixel $i$. 

We derive empirical relationships between photon noise and fiber coupling efficiency by varying $\rho$ as a function of $I$-band magnitude and spectral type. Nominal LBT observing conditions for our simulations include an $I=10$, M4V star at an airmass of $\sec{z}=1.4$ with 30 minutes of integration time under $\theta=1.1$" seeing. The Strehl ratio is estimated according to the SOUL WFS upgrade which has taken place since the original demonstrator measurements \citep{Pinna2016}. Residual tip-tilt errors are modeled with $\sigma_{\rm tip/tilt} = 8.8$ mas and NCPA errors (Z2,Z3) equivalent to $\rho = 0.8$.

\begin{figure}
	\centering
	\includegraphics[width =3.1in]{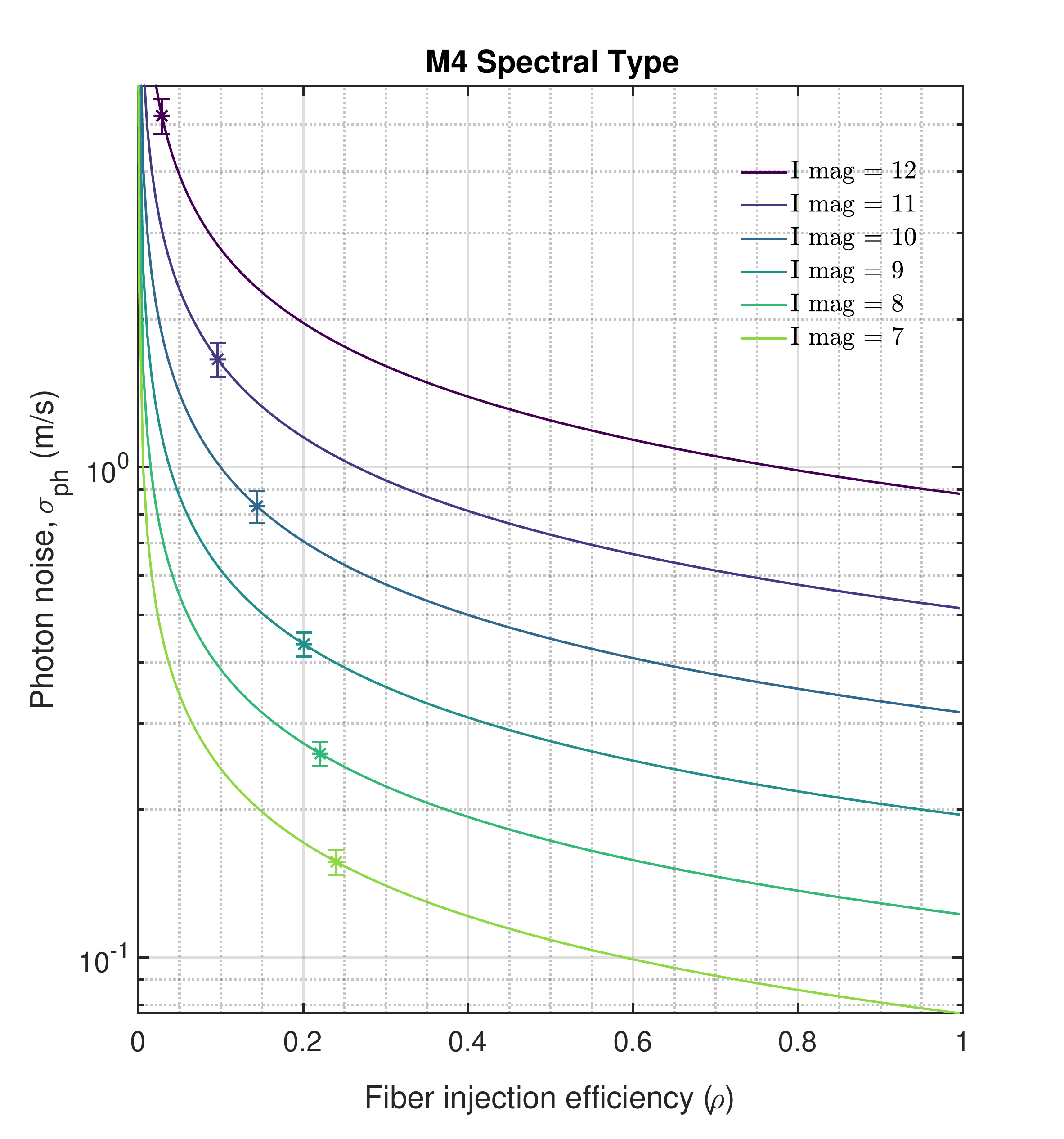}
	\caption{Photon noise ($\sigma_{\rm ph}$) as a function of fiber coupling efficiency ($\rho$) for M4V stars. Stars (*) on each line mark the expected value of $\rho$ for a given magnitude. $\rho$ is calculated using the nominal observational and conditions where $\rho_{\rm max} \approx 0.78$ based on LBT pupil geometry. Error bars indicate uncertainty in RV precision by based on variable fiber coupling conditions.}
	\label{fig:M4V_eng}
\end{figure}

Figure~\ref{fig:M4V_eng} shows photon noise limited RV precision as a function of fiber coupling for M4V stars. Smooth curves show the empirical relationship between fiber coupling and $\sigma_{\rm ph}$. An asterisk (*) marked on each line indicates expected values of $\rho$ for a given magnitude based on demonstrator results under nominal observing conditions. The uncertainty is obtained by varying fiber coupling coefficients according to results in previous sections.  

Empirical curves in Figure~\ref{fig:M4V_eng} follow the expectation that $\rho$ is proportional to the number of photons arriving ($N$), where $\sigma_{\rm ph} \propto 1/\sqrt{N}$. This demonstrates variations in fiber coupling efficiency improves RV precision in a similar fashion to simply increasing the total number of photons uniformly across the spectral band. However it is important to note only the single-measurement photon noise is accounted for in these simulations. In the presence of variable fiber alignment from one observation to the next, changes in chromatic fiber coupling efficiency introduce a systematic RV uncertainty \citep{Bechter2018}. To fully asses the impact of fiber coupling variations, these systematic RV uncertainties must be taken into account. 

Given assumptions for AO performance and coupling degradation found in the demonstrator system, coupling efficiencies slightly above 20\% per LBT dish are expected for $I = 7-9$ M-stars. At these magnitudes, simulations predicts $\sigma_{\rm ph} \leq 0.5m/s$. In this region, improvements in fiber coupling on the order of 5-10\% offer a few cm/s improvement in RV precision. However, improving $\rho$ by a similar percent can dramatically improve $\sigma_{\rm ph}$ for faint stars ($I\approx10-12$). 

It should be emphasized that the RV precision modeled in Figure~\ref{fig:M4V_eng} represents the theoretical photon noise floor of an $R\approx$200k SMF instrument with: a simultaneous bandpass of (0.970-1.28)$\mu m$, 30 minutes of integration time, and the dual 8.4m apertures of the LBT. Additional RV effects such as calibration errors and stellar activity are not included, which can increase the noise floor and degrade the true single-measurement precision~\citep{Bechter2018}.

While current SMF coupling efficiency is lower than MMFs, the total photon noise floor for iLocater is compensated largely by the LBT primary mirrors. For comparison, we note the existing precision of CARMENES instrument, an $R<80k$, extremely precise RV spectrograph located at the 3.5m telescope of the Calar Alto Observatory. CARMENES has a visible light arm which routinely achieves an internal precision of 1.6 m/s~\citep{Quirrenbach2018}. The near infrared arm has also been shown to derive an internal precision less than 2m/s on an M3.5V star with a $I=7.1$ (Luyten's star), a precision near to the photon noise floor~\citep{Quirrenbach2018}. HARPS, a visible light spectrograph with $R=115k$ and 3.6m aperture, is located at La Silla Observatory in Chile and has a single measurement precision of $\approx0.8$m/s for RV-quiet stars with SNR=200~\citep{Fischer2016}. Integrating longer can improve single measurement precision.

Although achieving efficient SMF injection presents a challenge, benefits of the SMF spectrograph are primarily in the calibration process and resolution. Unlike MMFs, variations in the illumination of SMFs do not result in focal ratio degradation nor modal noise; rather the fiber's output distribution is stable and decoupled from variations in the input beam~\citep{Lemke2010,McCoy2012,Schwab2012}. Simultaneously, the diffraction limited fiber core allows for high spectral resolution while using a compact diffraction grating. This is important as resolutions of $R\geq150k$ are needed to measure and possibly compensate for stellar activity~\citep{Davis2017}. The combination of stable spectrograph illumination and high resolution therefore provides a promising alternative method to multi-modal spectrograph optical designs.

\section{Conclusions}

On-sky AO-fed SMF coupling efficiency in the Y-band has been demonstrated at the 10-25\% level (absolute) at the LBT. Physical effects degrading coupling efficiency are separated into four coefficients: $\rho_{tel}$, $\rho_{strehl}$, $\rho_{tip/tilt}$, and $\rho_{ncpa}$. Telescope pupil geometry limits the maximum coupling to 78\%. The instantaneous Strehl ratio was shown to limits the maximum fiber coupling to 20-46\% and is strongly correlated with SMF coupling. Our calculations indicate that the demonstrator instrument coupled approximately 60\% of the available starlight after accounting for AO correction. 

Further degradation from the theoretical maximum coupling efficiency was caused by residual tip-tilt motion and static aberrations. Tip-tilt residuals have similar vibration signatures as the telescope structure, and degraded coupling by 9-23\%. Beam stabilization in the 0-30 Hz and 50-70 Hz bands will improve overall throughput and reduce noise cause by image jitter in fiber alignment. Static aberrations and fiber alignment combined resulted in an additional 20\% loss. The principal of measuring NCPA has been demonstrated through use of MGS phase retrieval, which can improve AO system offsets and boost the Strehl ratio in the fiber channel. A summary of fiber coupling coefficients can be found in Table~\ref{tab:conc}.

\begin{deluxetable}{cccccc}[h]
	\tablecaption{Summary}
	\tablehead{
	\colhead{$\rho_{\rm tel}$} & \colhead{$\rho_{\rm Strehl}$} & \colhead{$\rho_{\rm tip_tilt}$} & \colhead{$\rho_{\rm ncpa}$} & \colhead{$\rho_{\rm total}$}}
	\startdata
	$0.78$ & $0.33\pm0.13$ & $0.84\pm0.7$ & $\approx0.80$ & $0.18\pm0.7$
	\enddata
	\tablecomments{Fiber coupling terms, $\rho$, including telescope pupil, $tel$, the Strehl ratio, $strehl$, tip-tilt residuals, $tip/tilt$, and static aberrations $ncpa$}
	\label{tab:conc}
\end{deluxetable}
The simulated photon-noise-limited RV precision show promise to achieve $\sigma_{\rm ph} \leq 1$ m/s for bright stars ($I=7-10$) and $\sigma_{\rm ph}~1-10$ m/s for fainter magnitudes ($I=11,12$). Modest improvements in fiber coupling (5-10\%) may improve RV precision of faint stars at the m/s level and bright stars at the cm/s level. Additional consideration should be given to the impact of variable chromatic coupling efficiency when testing the repeatably of beam alignment and beam stability methods. 

\acknowledgements
The iLocater team would like to thank the LBTI team, especially Philip Hinz and Amali Vaz, for support operating the adaptive optics system. We are grateful for the help of LBTO staff and the INAF V-SHARK team, especially Fernando Pedichini, for assistance setting up fiber coupling experiments with LBTI. The LBT is an international collaboration among institutions in the United States, Italy and Germany. LBT Corporation partners are: The University of Arizona on behalf of the Arizona university system; Istituto Nazionale di Astrofisica, Italy; LBT Beteiligungsgesellschaft, Germany, representing the Max-Planck Society, the Astrophysical Institute Potsdam, and Heidelberg University; The Ohio State University, and The Research Corporation, on behalf of The University of Notre Dame, University of Minnesota and University of Virginia.

JRC acknowledges support from the NASA Early Career Fellowship and NSF CAREER Fellowship programs. The iLocater team is also grateful for contributions from the Potenziani family and the Wolfe family for their vision and generosity. 

A portion of this research was carried out at the Jet Propulsion Laboratory, California Institute of Technology, under a contract with the National Aeronautics and Space Administration.

\appendix

\section{Characterization Details}
\label{sec:deets}
\textcolor{white}{.} 

\begin{deluxetable*}{ccccccccccc} [h!]
\label{tab:deets}
    \tablecaption{Analysis Summary}
	\tablehead{
	\colhead{ID} & \colhead{Set \#} & \colhead{AO modes} &\colhead{Fiber Coupling} & \colhead{Strehl} & \colhead{$\sigma$(mas)} & \colhead{$\delta$ (mas/min)} &\colhead{Time (s)} & \colhead{fps (Hz)}  & \colhead{$\gamma_{sr}$}  & \colhead{$\gamma_{tip/tilt}$}}
	\startdata
	HD 89758 & I   & 300 & 0.14 $\pm0.02$ & 0.19 $\pm0.03$ & 8.8  & 0.29 & 190 & 123.8 & * & *\\ 
	HD 89758 & II  & 300 & 0.12 $\pm0.02$ & 0.18 $\pm0.03$ & 8.2  & 0.91 & 80  & 123.8 & 0.88 & 0.68\\ 
	HD 89758 & III & 300 & 0.09 $\pm0.02$ & 0.14 $\pm0.02$ & 10.2 & 2.76 & 80  & 123.8 & 0.65 & 0.63\\ 
	HD 89758 & IV  & 300 & 0.11 $\pm0.02$ & 0.16 $\pm0.03$ & 9.2  & 0.17 & 80  & 123.8 & 0.83 & 0.41\\ 
	HD 89758 & V   & 300 & 0.11 $\pm0.03$ & 0.18 $\pm0.04$ & 10.5 & 0.11 & 800 & 123.8 & 0.77 & 0.56\\ 
	HD 89758 & VI  & 153 & 0.08 $\pm0.01$ & 0.20 $\pm0.02$ & 12.3 & 0.45 & 16k & 34.9  & 0.69 & 0.37\\ 
  	SA0 82686& I   & 300 & 0.22 $\pm0.02$ & 0.28 $\pm0.04$ & 8.1  & 0.29 & 90  & 130   & * & *\\
	SA0 82686& II  & 400 & 0.22 $\pm0.02$ & 0.28 $\pm0.02$ & 7.4  & 0.17 & 380 & 130   & 0.51 & 0.28\\
	HD 113496& I   & 400 & 0.22 $\pm0.01$ & 0.28 $\pm0.03$ & 4.8  & 1.49 & 55  & 58.4  & 0.72 & 0.33\\
	\hline
	Closed Dome DX&I& -   & - & 0.83$\pm0.01$              & 11.1 & 2.41 & 30  & 3.1k  & - & -\\
    Closed Dome SX&I& -   & - & 0.83$\pm0.01$             & 10.2 & 2.46 & 30  & 3.1k  & - & -\\
	\enddata
	\tablecomments{Fiber coupling, Strehl ratio, and centroid scatter statistics for on-sky targets and closed dome calibration measurements including correlation parameters. The set~\# distinguishes separate observations of the same target. $\gamma_{\rm Strehl}$ is the correlation coefficient between fiber coupling and Strehl. Not all observations included simultaneous measurements of Strehl ratio and fiber coupling. These sets are marked with an asterisk (*). Observations without simultaneous fiber coupling and image data report statistics on complete sets while those with substantial time overlap report results on only the overlapping time period. $\sigma_{tip/tilt}$ is the $1\sigma$ scatter over the full observation, and $\delta$ is the systemic centroid drift. Elapsed time is the approximate duration of each data set and fps is frame rate. $\gamma_{tip/tilt}$ is the correlation between fiber coupling and tip-tilt residuals.}
\end{deluxetable*}

\bibliography{DBib} 

\begin{thebibliography}{}
\expandafter\ifx\csname natexlab\endcsname\relax\def\natexlab#1{#1}\fi
\providecommand{\url}[1]{\href{#1}{#1}}

\bibitem[{Bechter {et~al.}(2018)Bechter, Bechter, Crepp, Crass, \&
  King}]{Bechter2018}
Bechter, A., Bechter, E., Crepp, J.~R., Crass, J., \& King, D. 2018, 248.
\newblock \url{http://dx.doi.org/10.1117/12.2313658}

\bibitem[{Bechter {et~al.}(2015)Bechter, Crass, Ketterer, Crepp, King, Zhao,
  Reynolds, Hinz, Brooks, \& Bechter}]{Bechter2015}
Bechter, A., Crass, J., Ketterer, R., {et~al.} 2015, in Techniques and
  Instrumentation for Detection of Exoplanets VII, ed. S.~Shaklan, Vol. 9605
  (International Society for Optics and Photonics), 96051U.
\newblock
  \url{http://proceedings.spiedigitallibrary.org/proceeding.aspx?doi=10.1117/12.2188426}

\bibitem[{Bechter {et~al.}(2016)Bechter, Crass, Ketterer, Crepp, Reynolds,
  Bechter, Hinz, Pedichini, Foley, Runburg, Onuma, Gaudi, Micela, Pagano, \&
  Woodward}]{Bechter2016}
Bechter, A., Crass, J., Ketterer, R., {et~al.} 2016, in Adaptive Optics Systems
  V, Vol. 9909, 99092X.
\newblock \url{http://dx.doi.org/10.1117/12.2233153}

\bibitem[{Bechter {et~al.}(2019)Bechter, Bechter, Crepp, Crass, \&
  King}]{Bechter2019}
Bechter, E.~B., Bechter, A.~J., Crepp, J.~R., Crass, J., \& King, D. 2019,
  Publications of the Astronomical Society of the Pacific, 131,
  arXiv:1812.01649.
\newblock \url{http://dx.doi.org/10.1088/1538-3873/aaf278}

\bibitem[{Bikkannavar {et~al.}(2010)Bikkannavar, Redding, Green, Basinger,
  Cohen, Lou, Ohara, \& Shi}]{Bikkannavar2010}
Bikkannavar, S., Redding, D., Green, J., {et~al.} 2010, in Modern Technologies
  in Space- and Ground-based Telescopes and Instrumentation, Vol. 7739
  (International Society for Optics and Photonics), 77392X.
\newblock
  \url{http://proceedings.spiedigitallibrary.org/proceeding.aspx?doi=10.1117/12.858980}

\bibitem[{Blake {et~al.}(2015)Blake, Johnson, Plavchan, Sliski, Wittenmyer,
  Eastman, \& Barnes}]{Blake2015}
Blake, C., Johnson, J., Plavchan, P., {et~al.} 2015, American Astronomical
  Society, AAS Meeting {\#}225, id.257.32, 225.
\newblock \url{http://adsabs.harvard.edu/abs/2015AAS...22525732B}

\bibitem[{Bland-Hawthorn {et~al.}(2010)Bland-Hawthorn, Lawrence, Robertson,
  Campbell, Pope, Betters, Leon-Saval, Birks, Haynes, Cvetojevic, \&
  Jovanovic}]{Bland-Hawthorn2010}
Bland-Hawthorn, J., Lawrence, J., Robertson, G., {et~al.} 2010, in Ground-based
  and Airborne Instrumentation for Astronomy III, ed. I.~S. McLean, S.~K.
  Ramsay, \& H.~Takami, Vol. 7735 (International Society for Optics and
  Photonics), 77350N.
\newblock
  \url{http://proceedings.spiedigitallibrary.org/proceeding.aspx?doi=10.1117/12.856347}

\bibitem[{Burruss {et~al.}(2010)Burruss, Serabyn, Mawet, Roberts, Hickey,
  Rykoski, Bikkannavar, \& Crepp}]{Burruss2010}
Burruss, R.~S., Serabyn, E., Mawet, D.~P., {et~al.} 2010, in Adaptive Optics
  Systems II, ed. B.~L. Ellerbroek, M.~Hart, N.~Hubin, \& P.~L. Wizinowich,
  Vol. 7736 (International Society for Optics and Photonics), 77365X.
\newblock
  \url{http://proceedings.spiedigitallibrary.org/proceeding.aspx?doi=10.1117/12.857544}

\bibitem[{Butler {et~al.}(1996)Butler, Marcy, Williams, McCarthy, Dosanjh, \&
  Vogt}]{Butler1996}
Butler, R.~P., Marcy, G.~W., Williams, E., {et~al.} 1996, Publications of the
  Astronomical Society of the Pacific, 108, 500.
\newblock \url{http://iopscience.iop.org/article/10.1086/133755}

\bibitem[{Ciardi {et~al.}(2015)Ciardi, Beichman, Horch, \& Howell}]{Ciardi2015}
Ciardi, D.~R., Beichman, C.~A., Horch, E.~P., \& Howell, S.~B. 2015,
  Astrophysical Journal, 805, 16.
\newblock
  \url{http://stacks.iop.org/0004-637X/805/i=1/a=16?key=crossref.9eae8de02b7071fdcd11f349eb61b897}

\bibitem[{Crepp(2014)}]{Crepp2014}
Crepp, J.~R. 2014, Science, 346, 809.
\newblock \url{http://dx.doi.org/10.1126/science.1262071}

\bibitem[{Crepp {et~al.}(2016)Crepp, Crass, King, Bechter, Bechter, Ketterer,
  Reynolds, Hinz, Kopon, Cavalieri, Fantano, Koca, Onuma, Stapelfeldt, Thomes,
  Wall, Macenka, McGuire, Korniski, Zugby, Eisner, Gaudi, Hearty, Kratter,
  Kuchner, Micela, Nelson, Pagano, Quirrenbach, Schwab, Skrutskie, Sozzetti,
  Woodward, \& Zhao}]{Crepp2016}
Crepp, J.~R., Crass, J., King, D., {et~al.} 2016, in Ground-based and Airborne
  Instrumentation for Astronomy VI, ed. C.~J. Evans, L.~Simard, \& H.~Takami,
  Vol. 9908 (International Society for Optics and Photonics), 990819.
\newblock
  \url{http://proceedings.spiedigitallibrary.org/proceeding.aspx?doi=10.1117/12.2233135}

\bibitem[{Davis {et~al.}(2017)Davis, Cisewski, Dumusque, Fischer, \&
  Ford}]{Davis2017}
Davis, A.~B., Cisewski, J., Dumusque, X., Fischer, D.~A., \& Ford, E.~B. 2017,
  The Astrophysical Journal, 846, 59.
\newblock \url{http://dx.doi.org/10.3847/1538-4357/aa8303}

\bibitem[{Dekany {et~al.}(2013)Dekany, Roberts, Burruss, Bouchez, Truong,
  Baranec, Guiwits, Hale, Angione, Trinh, Zolkower, Shelton, Palmer, Henning,
  Croner, Troy, McKenna, Tesch, Hildebrandt, \& Milburn}]{Dekany2013}
Dekany, R., Roberts, J., Burruss, R., {et~al.} 2013, Astrophysical Journal,
  776, 130.
\newblock
  \url{http://stacks.iop.org/0004-637X/776/i=2/a=130?key=crossref.6cdf82f375a71c9c2c63a25f2d294a25}

\bibitem[{Esc{\'{a}}rate {et~al.}(2018)Esc{\'{a}}rate, Christou, Rahmer, Hill,
  Miller, \& Taylor}]{Escarate2018}
Esc{\'{a}}rate, P., Christou, J.~C., Rahmer, G., {et~al.} 2018, in Adaptive
  Optics Systems VI, ed. D.~Schmidt, L.~Schreiber, \& L.~M. Close, Vol. 10703
  (SPIE), 166.
\newblock
  \url{https://www.spiedigitallibrary.org/conference-proceedings-of-spie/10703/2313882/Vibration-environment-of-the-LBTOAO-system/10.1117/12.2313882.full}

\bibitem[{Esposito {et~al.}(2010)Esposito, Riccardi, Fini, Puglisi, Pinna,
  Xompero, Briguglio, Quir{\'{o}}s-Pacheco, Stefanini, Guerra, Busoni, Tozzi,
  Pieralli, Agapito, Brusa-Zappellini, Demers, Brynnel, Arcidiacono, \&
  Salinari}]{Esposito2010}
Esposito, S., Riccardi, A., Fini, L., {et~al.} 2010, in Adaptive Optics Systems
  II, ed. B.~L. Ellerbroek, M.~Hart, N.~Hubin, \& P.~L. Wizinowich, Vol. 7736
  (International Society for Optics and Photonics), 773609.
\newblock
  \url{http://proceedings.spiedigitallibrary.org/proceeding.aspx?doi=10.1117/12.858194}

\bibitem[{Fischer {et~al.}(2016)Fischer, Anglada-Escude, Arriagada, Baluev,
  Bean, Bouchy, Buchhave, Carroll, Chakraborty, Crepp, Dawson, Diddams,
  Dumusque, Eastman, Endl, Figueira, Ford, Foreman-Mackey, Fournier,
  Fűr{\'{e}}sz, Gaudi, Gregory, Grundahl, Hatzes, H{\'{e}}brard, Herrero,
  Hogg, Howard, Johnson, Jorden, Jurgenson, Latham, Laughlin, Loredo, Lovis,
  Mahadevan, McCracken, Pepe, Perez, Phillips, Plavchan, Prato, Quirrenbach,
  Reiners, Robertson, Santos, Sawyer, Segransan, Sozzetti, Steinmetz,
  Szentgyorgyi, Udry, Valenti, Wang, Wittenmyer, \& Wright}]{Fischer2016}
Fischer, D.~A., Anglada-Escude, G., Arriagada, P., {et~al.} 2016, Publications
  of the Astronomical Society of the Pacific, 128, 66001.
\newblock \url{http://eprv.astro.yale.edu.1}

\bibitem[{Green {et~al.}(2003)Green, Redding, Shaklan, \& Basinger}]{Green2003}
Green, J.~J., Redding, D.~C., Shaklan, S.~B., \& Basinger, S.~A. 2003, in
  High-Contrast Imaging for Exo-Planet Detection, ed. A.~B. Schultz \& R.~G.
  Lyon, Vol. 4860 (International Society for Optics and Photonics), 266.
\newblock
  \url{http://proceedings.spiedigitallibrary.org/proceeding.aspx?doi=10.1117/12.457883}

\bibitem[{Guyon(2003)}]{Guyon2003}
Guyon, O. 2003, Astronomy and Astrophysics, 404, 379.
\newblock \url{http://dx.doi.org/10.1051/0004-6361:20030457}

\bibitem[{Halverson {et~al.}(2015)Halverson, Roy, Mahadevan, Ramsey, Levi,
  Schwab, Hearty, \& Macdonald}]{Halverson2015}
Halverson, S., Roy, A., Mahadevan, S., {et~al.} 2015, Astrophysical Journal,
  806, arXiv:1505.07463v1.
\newblock \url{http://www.edmundoptics.com/}

\bibitem[{Harris {et~al.}(2018)Harris, Tepper, Davenport, Pedretti, Haynes,
  Hottinger, Anagnos, Nayak, Alonso, Deka, Minardi, Quirrenbach, Labadie, \&
  Haynes}]{Harris2018}
Harris, R.~J., Tepper, J., Davenport, J.~J., {et~al.} 2018, in Advances in
  Optical and Mechanical Technologies for Telescopes and Instrumentation III,
  ed. R.~Navarro \& R.~Geyl, Vol. 10706, International Society for Optics and
  Photonics (SPIE), 157 -- 171.
\newblock \url{https://doi.org/10.1117/12.2312009}

\bibitem[{Hippler(2018)}]{Hippler2018}
Hippler, S. 2018, Journal of Astronomical Instrumentation, 08, 1950001.
\newblock \url{http://arxiv.org/abs/1808.02693}

\bibitem[{Hottinger {et~al.}(2018)Hottinger, Harris, Dietrich, Blaicher,
  Gl{\"{u}}ck, Bechter, Pott, Sawodny, Quirrenbach, Crass, \&
  Koos}]{Hottinger2018}
Hottinger, P., Harris, R.~J., Dietrich, P.-I., {et~al.} 2018, in Advances in
  Optical and Mechanical Technologies for Telescopes and Instrumentation III,
  ed. R.~Geyl \& R.~Navarro, Vol. 10706 (SPIE), 77.
\newblock \url{http://dx.doi.org/10.1117/12.2312015}

\bibitem[{Jovanovic {et~al.}(2014)Jovanovic, Guyon, Martinache, Schwab, \&
  Cvetojevic}]{Jovanovic2014}
Jovanovic, N., Guyon, O., Martinache, F., Schwab, C., \& Cvetojevic, N. 2014,
  in Ground-based and Airborne Instrumentation for Astronomy V, ed. S.~K.
  Ramsay, I.~S. McLean, \& H.~Takami, Vol. 9147 (International Society for
  Optics and Photonics), 91477P.
\newblock
  \url{http://proceedings.spiedigitallibrary.org/proceeding.aspx?doi=10.1117/12.2057210}

\bibitem[{Jovanovic {et~al.}(2016{\natexlab{a}})Jovanovic, Schwab, Cvetojevic,
  Guyon, \& Martinache}]{Jovanovic2016}
Jovanovic, N., Schwab, C., Cvetojevic, N., Guyon, O., \& Martinache, F.
  2016{\natexlab{a}}, Publications of the Astronomical Society of the Pacific,
  128, arXiv:1609.06388.
\newblock \url{http://dx.doi.org/10.1088/1538-3873/128/970/121001}

\bibitem[{Jovanovic {et~al.}(2016{\natexlab{b}})Jovanovic, Cvetojevic, Schwab,
  Norris, Lozi, Gross, Betters, Singh, Guyon, Martinache, Doughty, \&
  Tuthill}]{Jovanovic2016a}
Jovanovic, N., Cvetojevic, N., Schwab, C., {et~al.} 2016{\natexlab{b}}, in
  Ground-based and Airborne Instrumentation for Astronomy VI, ed. C.~J. Evans,
  L.~Simard, \& H.~Takami, Vol. 9908 (International Society for Optics and
  Photonics), 99080R.
\newblock
  \url{http://proceedings.spiedigitallibrary.org/proceeding.aspx?doi=10.1117/12.2234299}

\bibitem[{Jovanovic {et~al.}(2017)Jovanovic, Schwab, Guyon, Lozi, Cvetojevic,
  Martinache, Leon-Saval, Norris, Gross, Doughty, Currie, \&
  Takato}]{Jovanovic2017}
Jovanovic, N., Schwab, C., Guyon, O., {et~al.} 2017, Astronomy and
  Astrophysics, 604, arXiv:1706.08821.
\newblock \url{http://dx.doi.org/10.1051/0004-6361/201630351}

\bibitem[{Lemke {et~al.}(2010)Lemke, Corbett, Allington-Smith, \&
  Murray}]{Lemke2010}
Lemke, U., Corbett, J., Allington-Smith, J., \& Murray, G. 2010, Modern
  Technologies in Space- and Ground-based Telescopes and Instrumentation, 7739,
  773924

\bibitem[{Mawet {et~al.}(2017)Mawet, Ruane, Xuan, Echeverri, Klimovich,
  Randolph, Fucik, Wallace, Wang, Vasisht, Dekany, Mennesson, Choquet, Delorme,
  \& Serabyn}]{Mawet2017}
Mawet, D., Ruane, G., Xuan, W., {et~al.} 2017, The Astrophysical Journal, 838,
  92.
\newblock
  \url{http://stacks.iop.org/0004-637X/838/i=2/a=92?key=crossref.b214f694afac6e726e521ecbcbf6f1b6}

\bibitem[{McCoy {et~al.}(2012)McCoy, Ramsey, Mahadevan, Halverson, \&
  Redman}]{McCoy2012}
McCoy, K.~S., Ramsey, L., Mahadevan, S., Halverson, S., \& Redman, S.~L. 2012,
  in Ground-based and Airborne Instrumentation for Astronomy IV, ed. I.~S.
  McLean, S.~K. Ramsay, \& H.~Takami, Vol. 8446, International Society for
  Optics and Photonics (SPIE), 1161 -- 1168.
\newblock \url{https://doi.org/10.1117/12.926287}

\bibitem[{Noll(1976)}]{Noll1976}
Noll, R.~J. 1976, J Opt Soc Am, 66, 207.
\newblock \url{https://www.osapublishing.org/abstract.cfm?URI=josa-66-3-207}

\bibitem[{Pinna {et~al.}(2016)Pinna, Esposito, Hinz, Agapito, Bonaglia,
  Puglisi, Xompero, Riccardi, Briguglio, Arcidiacono, Carbonaro, Fini, Montoya,
  \& Durney}]{Pinna2016}
Pinna, E., Esposito, S., Hinz, P., {et~al.} 2016, in Adaptive Optics Systems V,
  ed. E.~Marchetti, L.~M. Close, \& J.-P. V{\'{e}}ran, Vol. 9909 (International
  Society for Optics and Photonics), 99093V.
\newblock
  \url{http://proceedings.spiedigitallibrary.org/proceeding.aspx?doi=10.1117/12.2234444}

\bibitem[{Quirrenbach {et~al.}(2018)Quirrenbach, Amado, Ribas, Caballero,
  Seifert, Aceituno, Azzaro, Barrado, Becerril, B{\`{e}}jar, Ben{\'{i}}tez,
  Brinkm{\"{o}}ller, Colom{\'{e}}, Cort{\'{e}}s-Contreras, Czesla,
  Fr{\"{o}}lich, Galad{\'{i}}-Enr{\'{i}}quez, {Gonz{\'{a}}lez Hern{\'{a}}ndez},
  {Gonz{\'{a}}lez Peinado}, Guenther, de~Guindos, Hagen, Henning,
  {Hern{\'{a}}ndez Casta{\~{n}}o}, Herrero, Hintz, Jeffers, Kaminski, Klahr,
  Marfil, Mart{\'{i}}n, Mart{\'{i}}n-Ruiz, Mathar, Montes, Morales, Nagel,
  Pall{\'{e}}, P{\'{e}}rez-Medialdea, Perger, Rebolo, Reffert, Rosich, Sabotta,
  Sch{\"{a}}fer, Schiller, Schweitzer, Solano, Stahl, {Tala Pinto}, Trifonov,
  Yan, Zechmeister, Abell{\'{a}}n, Abril, Alonso-Floriano, {Ammler-von Eiff},
  Anglada-Escud{\'{e}}, Anwand-Heerwart, Berdi{\~{n}}as, Bergondy, del Burgo,
  C{\'{a}}rdenas, Casal, Claret, Ferro, G{\'{a}}lvez-Ortiz, Gesa, {G{\'{o}}mez
  Galera}, Guijarro, Hedrosa, Hermann, Hermelo, {Hern{\'{a}}ndez Arab{\'{i}}},
  Hidalgo, Huber, Huber, Kehr, Klein, Kl{\"{u}}ter, Klutsch, Labarga, Labiche,
  Lamert, Lemke, Lenzen, Lizon, Lodieu, L{\'{o}}pez-Morales, {L{\'{o}}pez
  Salas}, L{\'{o}}pez-Santiago, Mart{\'{i}}nez-Rodr{\'{i}}guez, {Maroto
  Fern{\'{a}}ndez}, Marvin, Mirabet, Moreno-Raya, Moya, Naranjo, Pascual,
  P{\'{e}}rez-Calpena, Perryman, Rohloff, {S{\'{a}}nchez Carrasco}, Schmidt,
  Strachan, Tal-Or, Tulloch, Veredas, Vilardell, Wagner, Zhao, Reiners, Baroch,
  Bauer, {Cardona Guill{\'{e}}n}, Cifuentes, Dreizler, Fuhrmeister, Hatzes,
  Hauschildt, Helmling, Herbort, Johnson, de~Juan, K{\"{u}}rster, Lafarga,
  Sairam, Lamp{\'{o}}n, Lara, Launhardt, {L{\'{o}}pez del Fresno},
  L{\'{o}}pez-Puertas, Luque, Mandel, Nortmann, Nowak, Passegger, Pavlov,
  Pedraz, Rodr{\'{i}}guez, {Rodr{\'{i}}guez L{\'{o}}pez}, Sadegi, Salz,
  S{\'{a}}nchez-L{\'{o}}pez, Sanz-Forcada, Sarkis, Schmitt, Sch{\"{o}}fer,
  Shulyak, {Zapatero Osorio}, Arroyo-Torres, Bl{\"{u}}mcke, Cano, Carro,
  D{\'{i}}ez-Alonso, Doellinger, Dorda, Feiz, Fern{\'{a}}ndez, Gaisn{\'{e}},
  Gallardo, Garc{\'{i}}a-Piquer, Garc{\'{i}}a-Vargas, Garrido,
  Gonz{\'{a}}lez-{\'{A}}lvarez, Gonz{\'{a}}lez-Cuesta, Grohnert,
  Gr{\"{o}}zinger, Gu{\`{a}}rdia, {Hern{\'{a}}ndez Hernando}, Holgado, Huke,
  Kim, Laun, L{\'{a}}zaro, Llamas, {L{\'{o}}pez Gonz{\'{a}}lez}, {Mag{\'{a}}n
  Madinabeitia}, Mall, Mancini, {Mar{\'{i}}n Molina}, Mundt, Panduro, Pluto,
  Ram{\'{o}}n, Redondo, Reinhart, Rhode, Rix, Rodler, S{\'{a}}nchez-Blanco,
  Sarmiento, Storz, St{\"{u}}rmer, Su{\'{a}}rez, Tabernero, Ulbrich, {Vico
  Linares}, Vidal-Dasilva, Winkler, Wolthoff, \& Xu}]{Quirrenbach2018}
Quirrenbach, A., Amado, P.~J., Ribas, I., {et~al.} 2018, in Ground-based and
  Airborne Instrumentation for Astronomy VII, ed. H.~Takami, C.~J. Evans, \&
  L.~Simard, Vol. 10702 (SPIE), 32.
\newblock
  \url{https://www.spiedigitallibrary.org/conference-proceedings-of-spie/10702/2313689/CARMENES--high-resolution-spectra-and-precise-radial-velocities-in/10.1117/12.2313689.full}

\bibitem[{{Roberts, Jr.} {et~al.}(2004){Roberts, Jr.}, Perrin, Marchis,
  Sivaramakrishnan, Makidon, Christou, Macintosh, Poyneer, van Dam, \&
  Troy}]{RobertsJr.2004}
{Roberts, Jr.}, L.~C., Perrin, M.~D., Marchis, F., {et~al.} 2004, in
  Advancements in Adaptive Optics, Vol. 5490 (International Society for Optics
  and Photonics), 504.
\newblock
  \url{http://proceedings.spiedigitallibrary.org/proceeding.aspx?doi=10.1117/12.549115}

\bibitem[{Robertson \& Bland-Hawthorn(2012)}]{Robertson2012}
Robertson, J.~G., \& Bland-Hawthorn, J. 2012, Ground-based and Airborne
  Instrumentation for Astronomy IV, 8446, 844623.
\newblock \url{http://dx.doi.org/10.1117/12.924937}

\bibitem[{Ross(2009)}]{Ross2009}
Ross, T.~S. 2009, Applied Optics, 48, 1812.
\newblock \url{https://www.osapublishing.org/abstract.cfm?URI=ao-48-10-1812}

\bibitem[{Ruilier \& Cassaing(2001)}]{Ruilier2001}
Ruilier, C., \& Cassaing, F. 2001, Journal of the Optical Society of America A,
  18, 143.
\newblock \url{https://www.osapublishing.org/abstract.cfm?URI=josaa-18-1-143}

\bibitem[{Schwab {et~al.}(2012)Schwab, Leon-Saval, Betters, Bland-Hawthorn, \&
  Mahadevan}]{Schwab2012}
Schwab, C., Leon-Saval, S.~G., Betters, C.~H., Bland-Hawthorn, J., \&
  Mahadevan, S. 2012in , 403--406.
\newblock \url{http://arxiv.org/abs/1212.4867}

\bibitem[{Serabyn {et~al.}(2010)Serabyn, Mennesson, Martin, Liewer, Mawet,
  Hanot, Loya, Colavita, \& Ragland}]{Serabyn2010}
Serabyn, E., Mennesson, B., Martin, S., {et~al.} 2010, in Optical and Infrared
  Interferometry II, Vol. 7734 (International Society for Optics and
  Photonics), 77341E.
\newblock
  \url{http://proceedings.spiedigitallibrary.org/proceeding.aspx?doi=10.1117/12.857753}

\bibitem[{Shaklan \& Roddier(1988)}]{Shaklan1988}
Shaklan, S., \& Roddier, F. 1988, Applied Optics, 27, 2334.
\newblock \url{https://www.osapublishing.org/abstract.cfm?URI=ao-27-11-2334}

\bibitem[{Sivaramakrishnan {et~al.}(2001)Sivaramakrishnan, Koresko, Makidon,
  Berkefeld, \& Kuchner}]{Sivaramakrishnan2001}
Sivaramakrishnan, A., Koresko, C.~D., Makidon, R.~B., Berkefeld, T., \&
  Kuchner, M.~J. 2001, The Astrophysical Journal, 552, 397.
\newblock \url{http://stacks.iop.org/0004-637X/552/i=1/a=397}

\bibitem[{Snyder \& Love(1983)}]{Snyder1983}
Snyder, A.~W., \& Love, J. D. J.~D. 1983, {Optical waveguide theory} (Chapman
  and Hall), 734

\bibitem[{Wagner \& Tomlinson(1982)}]{Wagner1982}
Wagner, R.~E., \& Tomlinson, W.~J. 1982, Applied Optics, 21, 2671.
\newblock \url{https://www.osapublishing.org/abstract.cfm?URI=ao-21-15-2671}

\bibitem[{Wang {et~al.}(2017)Wang, Mawet, Ruane, Hu, \& Benneke}]{Wang2017}
Wang, J., Mawet, D., Ruane, G., Hu, R., \& Benneke, B. 2017, The Astronomical
  Journal, 153, 183.
\newblock
  \url{http://stacks.iop.org/1538-3881/153/i=4/a=183?key=crossref.4d38296cac6f166844e2fcdaaab6494b}

\end{thebibliography}

\end{document}